\pdfoutput=1

\documentclass[aps,prd,amsmath,superscriptaddress,%
               nobibnotes,nofootinbib,preprintnumbers,
               onecolumn,12pt,tightenlines]{revtex4}

\usepackage{graphicx}

\usepackage{ifthen}

\graphicspath{{figures/}}

\newcommand{\ie}{i.e.}
\newcommand{\eg}{e.g.}

\newcommand{\refeqn}[2][eqn:]{Eqn.~(\ref{#1#2})}

\newcommand{\reftab}[2][tab:]{Table~\ref{#1#2}}

\newcommand{\reffig}[2][fig:]{Figure~\ref{#1#2}}
\newcommand{\Reffig}[2][fig:]{Figure~\ref{#1#2}}
\newcommand{\refsec}[2][sec:]{Section~\ref{#1#2}} 

\newcommand{\ifmulticol}[2]{%
  \ifthenelse{\lengthtest{1.9\columnwidth<\textwidth}}{#1}{#2}%
}

\newcommand{\insertfig}[2][\scfigwidth]{%
    \hspace*{\stretch{1}}
    \includegraphics[keepaspectratio,width=#1\columnwidth]{#2}
    \hspace*{\stretch{1}}
}

\newcommand{\orderof}[1]{\ensuremath{\mathcal{O}(#1)}}

\newcommand{\gae}{%
  \ensuremath{\lower 2pt \hbox{%
    $\, \buildrel {\scriptstyle >}\over {\scriptstyle \sim}\,$}%
    }%
  }
\newcommand{\lae}{%
  \ensuremath{\lower 2pt \hbox{%
    $\, \buildrel {\scriptstyle <}\over {\scriptstyle \sim}\,$}%
    }%
  }

\newcommand{\scinot}[2]{\ensuremath{#1\times\!\!10^{#2}}}

\newcommand{\dof}{\textit{dof}}

\newcommand{\chimin}{\ensuremath{\chi_{\mathrm{min}}}}

\newcommand{\bsigma}{\ensuremath{\boldsymbol{\sigma}}}

\newcommand{\Enr}{\ensuremath{E_{\mathrm{nr}}}}
\newcommand{\Eee}{\ensuremath{E_{\mathrm{ee}}}}

\newcommand{\dRdEnr}{\ensuremath{\frac{dR}{dE_{\mathrm{nr}\!\!\!\!\!}}\;}}
\newcommand{\dRdEee}{\ensuremath{\frac{dR}{dE_{\mathrm{ee}\!\!\!\!\!}}\;}}

\newcommand{\mchi}{\ensuremath{m_{\chi}}}
\newcommand{\rhochi}{\ensuremath{\rho_{\chi}}}
\newcommand{\nchi}{\ensuremath{n_{\chi}}}

\newcommand{\vmin}{\ensuremath{v_\mathrm{min}}}

\newcommand{\vesc}{\ensuremath{v_\mathrm{esc}}}

\newcommand{\bu}{\ensuremath{\mathbf{u}}}  
\newcommand{\bv}{\ensuremath{\mathbf{v}}}  

\newcommand{\SHM}{\ensuremath{\mathrm{SHM}}}   

\newcommand{\qmax}{\ensuremath{q_{\mathrm{max}}}}  

\newcommand{\fpSI}{\ensuremath{f_{\mathrm{p}}}}
\newcommand{\fnSI}{\ensuremath{f_{\mathrm{n}}}}
\newcommand{\apSD}{\ensuremath{a_{\mathrm{p}}}}
\newcommand{\anSD}{\ensuremath{a_{\mathrm{n}}}}

\newcommand{\sigmaSI}{\ensuremath{\sigma_{\mathrm{SI}}}}
\newcommand{\sigmaSD}{\ensuremath{\sigma_{\mathrm{SD}}}}

\newcommand{\sigmapSI}{\ensuremath{\sigma_{\mathrm{p,SI}}}}

\newcommand{\sigmapSD}{\ensuremath{\sigma_{\mathrm{p,SD}}}}
\newcommand{\sigmanSD}{\ensuremath{\sigma_{\mathrm{n,SD}}}}

\newcommand{\Sp}{\ensuremath{\langle S_{\mathrm{p}} \rangle}}
\newcommand{\Sn}{\ensuremath{\langle S_{\mathrm{n}} \rangle}}

\begin{document}


\preprint{MCTP-13-14}
\preprint{NSF-KITP-13-104}
\preprint{CETUP2013-004}

\title{Lowering the Threshold in the DAMA Dark Matter Search}

\author{Chris Kelso}
\email[]{ckelso@physics.utah.edu}
\affiliation{
 Department of Physics \& Astronomy,
 University of Utah,
 Salt Lake City, UT 84112}

\author{Pearl Sandick}
\email[]{sandick@physics.utah.edu}
\affiliation{
 Department of Physics \& Astronomy,
 University of Utah,
 Salt Lake City, UT 84112}

\author{Christopher Savage}
\email[]{savage@physics.utah.edu}
\affiliation{
 Department of Physics \& Astronomy,
 University of Utah,
 Salt Lake City, UT 84112}

\date{\today}



\begin{abstract} 

\vspace{0.1in}

The DAMA experiment searches for Weakly Interacting Massive Particle
(WIMP) dark matter via its expected but rare interactions within the
detector, where the interaction rates will modulate throughout the
year due to the orbital motion of the Earth.  Over the course of more
than 10 years of operation, DAMA has indeed detected a strong
modulation in the event rate above the detector threshold of 2~keVee.
Under standard assumptions regarding the dark matter halo and WIMP
interactions, this signal is consistent with that expected of WIMPs of
two different approximate masses: $\sim$10~GeV and $\sim$70~GeV.  We
examine how a lower threshold, allowed by recent upgrades to the DAMA
detector, may shed light on this situation.  We find that the lower
threshold data should rule out one of the two mass ranges for
spin-independent couplings (in the worst case, disfavoring one of the
masses by still more than 2.6$\sigma$) and is likely, though not
certain, to do the same for spin-dependent couplings.
Furthermore, the data may indicate whether the interaction is
predominantly spin-independent or spin-dependent in some cases.
Our findings illustrate the importance of a low threshold in modulation
searches.

\end{abstract} 

\maketitle



\section{\label{sec:Intro}Introduction}

Although there exists an abundance of evidence at many different
energy and length scales that the vast majority of matter in our
universe is non-baryonic and does not possess significant
electromagnetic interactions, the nature of this dark matter remains
unknown. Understanding the particle nature of dark matter is one of
the most important outstanding problems in both cosmology and particle
physics.  There are currently numerous direct detection experiments
searching for interactions between the dark matter particles and the
nuclei of the detectors.

The DAMA experiment is the first direct detection experiment to make a
claim of the detection of particle dark matter.  For more than a
decade, the DAMA/LIBRA collaboration (and its predecessor DAMA/NaI)
have claimed detection of an annual modulation that they associate
with dark matter interacting with the NaI crystals of their detectors
\cite{Bernabei:2003za,Bernabei:2008yh}.  The DAMA annual modulation is
currently reported as nearly a 9$\sigma$ effect
\cite{Bernabei:2010mq}, and is consistent with a $\sim$70 or
$\sim$10~GeV WIMP elastically scattering predominantly off of iodine
or sodium, respectively~\cite{Bottino:2003iu,Bottino:2003cz,
Gondolo:2005hh,Petriello:2008jj,Chang:2008xa,Fairbairn:2008gz,
Savage:2008er}.

Hints of light dark matter may now be evident at other experiments, as
well.  In February of 2010, the CoGeNT collaboration reported an
excess of scattering events at low energy~\cite{Aalseth:2010vx}. The
collaboration subsequently reported 2.8$\sigma$ evidence for an annual
modulation in these excess events~\cite{Aalseth:2011wp}.  At the same
conference, CRESST-II also announced events above their known
backgrounds that can be interpreted as evidence for light dark
matter~\cite{Angloher:2011uu}.  Whether DAMA, CoGeNT, and CRESST are
consistent in the low-mass window is still
debated~\cite{Kelso:2011gd,Fox:2011px}.

To complicate the matter even further, CDMS has not observed an annual
modulation in their germanium detectors~\cite{Ahmed:2012vq}, although
they have not yet released results with an energy threshold low enough
to be sensitive to the entire energy range for the DAMA modulation.
In addition, CDMS \cite{Ahmed:2009zw,Ahmed:2010wy} and both
XENON10~\cite{Angle:2011th} and XENON100~\cite{Aprile:2012nq} find
null results at low energies that appear to be in conflict with the
three experiments that report anomalies.

The very recent development of three events observed in the signal
window of the silicon detectors of CDMS~\cite{Agnese:2013rvf} has only
increased the interest (and confusion) in the low mass region.  The
CDMS result is even more intriguing given that the mass and cross
section consistent with the dark matter interpretation of the three
events lands in the same area of parameter space as that indicated by
the CoGeNT excess, accounting for the latter's updated estimates for
surface event contamination~\cite{Kelso:2011gd,Aalseth:2012if}.

The DAMA/LIBRA experiment has recently undergone a significant
upgrade, and has been taking data in the DAMA/LIBRA-phase~2
configuration since January 2011~\cite{Bernabei:2013qx}.  In November
of 2010, all of the photo-multiplier tubes (PMT's) in the DAMA/LIBRA
experiment were replaced by high quantum efficiency
PMT's~\cite{Bernabei:2012zzb}. One of the most interesting aspects of
this detector upgrade is the anticipated lowering of the energy
threshold from 2~keVee down to 1~keVee.  In this paper we will examine
the extent to which DAMA's lower energy threshold can help to clarify
the currently confusing situation for light dark matter.  We will
start by reviewing the basics of direct detection in
\refsec{Detection}, followed by a discussion of the DAMA experiment
and results in \refsec{DAMA}.  We examine implications of this lower
threshold in \refsec{Models} and summarize our findings in
\refsec{Conclusions}.

\section{\label{sec:Detection}Dark Matter Direct Detection}

Dark matter direct detection experiments aim to observe the recoil of
a nucleus in a collision with a dark matter
particle~\cite{Goodman:1984dc}.  After an elastic collision with a
WIMP $\chi$ of mass $\mchi$, a nucleus of mass $M$ recoils with energy
$\Enr = (\mu^2 v^2/M)(1-\cos\theta)$, where $\mu \equiv \mchi M/
(\mchi + M)$ is the reduced mass of the WIMP-nucleus system, $v$ is
the speed of the WIMP relative to the nucleus, and $\theta$ is the
scattering angle in the center of mass frame.  The differential recoil
rate per unit detector mass is
\begin{equation}\label{eqn:dRdEnr}
  \dRdEnr
    = \frac{\nchi}{M} \, \Big\langle v \frac{d\sigma}{d\Enr}  \Big\rangle
    = \frac{2\rhochi}{\mchi}
      \int d^3v \, v f(\bv,t) \frac{d\sigma}{dq^2}(q^2,v) \, ,
\end{equation}
where $\nchi = \rhochi/\mchi$ is the local number density of WIMPs and
$\rhochi$ is the local dark matter mass density, $f(\bv,t)$ is the
time-dependent WIMP velocity distribution, and
$\frac{d\sigma}{dq^2}(q^2,v)$ is the velocity-dependent differential
cross-section with the momentum exchange in the scatter given by $q^2
= 2 M \Enr$.  The differential rate is typically given in units of
cpd\,kg$^{-1}$\,keV$^{-1}$, where cpd is counts per day.  Below, we
briefly describe the particle physics and astrophysics terms of the
equation above.  More detailed reviews of the dark matter scattering
process and direct detection can be found in Refs.~\cite{Primack:1988zm,
Smith:1988kw,Lewin:1995rx,Jungman:1995df,Bertone:2004pz,Freese:2012xd}.


\textit{Particle physics.}
Most WIMP candidates have WIMP-quark couplings that give rise to
spin-independent (SI) and spin-dependent (SD) WIMP-nucleus interactions
with a differential cross-section of the form:
\begin{equation}\label{eqn:dsigmadq}
  \frac{d\sigma}{dq^2}(q^2,v)
    = \frac{\sigma_{0}}{4 \mu^2 v^2} F^2(q) \, \Theta(\qmax-q) \, .
\end{equation}
Here, $\Theta$ is the Heaviside step function, $\qmax = 2 \mu v$ is
the maximum momentum transfer in a collision at a relative velocity
$v$, $\sigma_0$ is the scattering cross-section in the
zero-momentum-transfer limit---we shall use $\sigmaSI$ and $\sigmaSD$
to represent this term in the SI and SD cases, respectively---and
$F^2(q)$ is a form factor to account for the finite size of the
nucleus.  The total WIMP-nucleus scattering rate is determined by
summing the SI and SD contributions, each with its own value of the
form factor.  A description of form factors can be found in
Refs.~\cite{Helm:1956zz,Lewin:1995rx} (SI) and
Refs.~\cite{Bednyakov:2004xq,Bednyakov:2006ux} (SD).

The WIMP-nucleus cross-section $\sigma_0$ can be given in terms of
WIMP-nucleon couplings, though with different scaling behaviors for
the two types of interactions.  In the SI case,
\begin{equation} \label{eqn:sigmaSI2}
  \sigmaSI = \frac{4}{\pi} \mu^2
             \Big[ Z \fpSI + (A-Z) \fnSI \Big]^{2} \; ,
\end{equation}
where $A$ is the atomic mass number, $Z$ is the number of protons, and
$\fpSI$ and $\fnSI$ are the effective couplings to protons and
neutrons, respectively.  The common assumption is that WIMPs couple to
protons and neutrons with nearly equal strength for SI interactions,
leading to an $A^2$ enhancement to the signal for heavier targets.
While isospin-violating interactions, where the SI WIMP-proton and
WIMP-neutron couplings can substantially differ, have also been
explored in the literature \cite{Kurylov:2003ra,Chang:2010yk,
Feng:2011vu}, we will not examine their effects in this paper.

For the SD case,
\begin{equation} \label{eqn:sigmaSD}
  \sigmaSD = \frac{32 \mu^2}{\pi} G_{F}^{2} J(J+1) \Lambda^2 \, ,
\end{equation}
where $G_F$ is the Fermi constant, $J$ is the spin of the nucleus, and
\begin{equation} \label{eqn:Lambda}
  \Lambda \equiv \frac{1}{J} \Big( \apSD \Sp + \anSD \Sn \Big) \, ,
\end{equation}
where $\Sp$ and $\Sn$ are the average spin contributions from the
proton and neutron groups, respectively, and $\apSD$ ($\anSD$) are the
effective couplings to the proton (neutron).  Unlike the SI case, the
WIMP-proton and WIMP-neutron couplings are not expected to be nearly
equal in the SD case.


\textit{Astrophysics.}
The velocity distribution, $f(\bv,t)$, and the local dark matter
density, $\rhochi$, contain all the necessary information about the
local distribution of dark matter to calculate the expected dark
matter signal in a detector. The most commonly employed assumption
when analyzing direct detection data is that dark matter is uniformly
distributed throughout an isothermal sphere.  The velocity
distribution for this Standard Halo Model (SHM)
\cite{Drukier:1986tm,Freese:1987wu} is
\begin{equation} \label{eqn:SHMDist}
  f_{\SHM}(\bu) =
    \begin{cases}
      \frac{1}{N v_0^3\pi^{3/2}} e^{-u^2/v_0^2} , 
        & \textrm{if} \,\,u < \vesc  \\
      0 , & \textrm{otherwise}
    \end{cases}
\end{equation}
for dark matter particles with velocity $\bu$ (in the halo rest
frame), Galactic escape velocity $\vesc$, and with $N$ being a
normalization constant that ensures $\int{d^3u\,f(\bu)} = 1$.  The
velocity dispersion, $v_0$, is the most likely speed for a dark matter
particle with this velocity distribution, which is equal to the local
circular speed of the sun in the SHM. This function should be thought
of as a reasonable, but approximate, parametrization of the dark
matter's true velocity distribution.

There still exists a significant amount of uncertainty in the
astrophysical parameters relevant to the SHM.  Recent estimates for
the local circular speed generally place it at 235~km/s
$\pm$10\%~\cite{Reid:2009nj,McMillan:2009yr,Bovy:2009dr}, somewhat
higher than the historical canonical value of
220~km/s~\cite{Kerr:1986hz}, though the latter value remains viable.
A sample of high velocity stars from the RAVE survey has been used to
place the Galactic escape velocity in the range of 498--608~km/s at
the 90\% confidence level (CL) with a median likelihood of
544~km/s~\cite{Smith:2006ym}.  While 0.3~GeV/cm$^3$ has long been
taken as the canonical value for the local density of the smooth dark
matter component, recent estimates tend to favor 0.4~GeV/cm$^3$,
though these estimates are model dependent and vary in the literature
by as much as a factor of two~\cite{Caldwell:1981rj,
  Catena:2009mf,Weber:2009pt,Salucci:2010qr,Pato:2010yq}.  In this
work, we will use $v_0=235$~km/s, $\vesc=550$~km/s and
$\rhochi=0.4$~GeV/cm$^3$.

The motion of the earth through the Galactic halo will induce a time
dependence in a dark matter signal in direct detection
experiments~\cite{Drukier:1986tm,Freese:1987wu}.  The integral over
the velocity distribution in \refeqn{dRdEnr} must be evaluated for the
SHM with $\bu=\bv_e+\bv$, where $\bu$ is the WIMP velocity (in the
halo's rest frame), $\bv_e$ is the velocity of the Earth (in the
halo's rest frame) and $\bv$ is the velocity of the WIMP relative to
the nucleus.  The Earth's speed relative to that of the Galactic halo
is given by $v_e=v_{\odot}+v_{\rm orb} \cos\gamma\,\cos[\omega(t-t_0)]$
where $v_{\odot}=v_0+13$~km/s is the speed of the sun relative to the
rest frame of the halo, the orbital speed of Earth about the sun is
$v_{\rm orb}=30$~km/s, $\cos\gamma=0.49$ accounts for the direction of
motion of the Earth relative to that of the halo, $t_0$ is the date at
which Earth moves fastest relative to the halo (around June~2), and
$\omega=2\pi/{\rm year}$.  Thus, in the SHM, the Earth's motion will
lead to a sinusoidal modulation of the signal in a dark matter
detector.

\section{\label{sec:DAMA}The DAMA Experiment}

In this section, we discuss the DAMA experimental setup and current
results.  We also address two issues affecting the interpretation of
these results: the choice of binning for the data and uncertainties in
the quenching factor used to calibrate the energy scale of events in
the DAMA detectors.

\subsection{\label{sec:overview}Overview}

The DAMA/LIBRA collaboration uses 25 highly radio-pure NaI(Tl)
detectors with a total target mass of $\sim$250 kg.  Each detector is
instrumented with PMT's that measure the scintillation light created
by a nuclear recoil; these PMT's have been recently
upgraded~\cite{Bernabei:2012zzb}.  The collaboration uses gamma ray
sources for calibrating their detectors that produce scintillation
light through {\it electron} recoils.  This means that the energy of a
nuclear recoil will be measured in electron equivalent energy, or
keVee.  The electron equivalent energy, $\Eee$, is related to the
nuclear recoil energy, $\Enr$, via what is termed the ``quenching
factor.'' For the DAMA detectors, the quenching factor, $Q$, is the
ratio of the amount of scintillation light created by a recoiling
nucleus of some kinetic energy to that created by a recoiling electron
of the same kinetic energy, \ie\ $\Eee = Q\Enr$.

Measuring the value of the quenching factor for sodium at the low
energies relevant for light WIMPs is quite difficult.  This has led to
some uncertainty as to the value and possible energy dependence of the
quenching factor (see Ref.~\cite{Collar:2013gu}).  In our work, we
adopt the values of the DAMA collaboration, $Q_{\rm Na}=0.3$ and
$Q_{\rm I}=0.09$ \cite{Bernabei:1996vj} (see
Ref.~\cite{Tretyak:2009sr} and references therein for quenching factor
measurements of NaI and several other scintillators used in direct
detection experiments). Broadly speaking, increasing the quenching
factor will shift an allowed region in the (mass, cross section) plane
towards lower masses while leaving the cross section values relatively
unaffected (see Ref.~\cite{Kelso:2011gd}).

Every real detector has both an energy resolution and a detection
efficiency for relevant events.  We implement these effects for the
DAMA/LIBRA detectors using
\begin{equation}\label{eqn:dRdEee}
  \dRdEee(\Eee,t)
    = \int_0^{\infty} d\Enr \,
        \varepsilon(Q\Enr) \, \phi(\Enr,\Eee) \, \dRdEnr(\Enr,t).
\end{equation}
Here, $\varepsilon(Q\Enr)$ is the event detection efficiency and
\begin{equation}\label{eqn:phi}
  \phi(\Enr,\Eee) = \frac{1}{\sqrt{2\pi\sigma^2(Q\Enr)}}
                    e^{-(\Eee-Q\Enr)^2/2\sigma^2(Q\Enr)}
\end{equation}
is the differential response function, defined such that
$\phi(\Enr,\Eee)\,\Delta\Eee$ is the probability that a nuclear recoil
of energy $\Enr$ will produce a scintillation signal measured between
$\Eee$ and $\Eee+\Delta\Eee$ (in the limit of small $\Delta\Eee$).
The energy resolution for the detectors with the original PMT's at the
energies of interest is given by~\cite{Bernabei:2008yh}
\begin{equation}\label{eqn:sigmaER}
  \sigma(Q\Enr) = \alpha\sqrt{Q\Enr}+\beta\,Q\Enr
\end{equation}
with $\alpha=(0.448\pm0.035)\sqrt{\rm keVee}$ and
$\beta=(9.1\pm5.1)\times 10^{-3}$.  The collaboration presents an
efficiency corrected spectrum, so we can safely use
$\varepsilon(Q\Enr)=1$.  Although most of the upgraded detectors do
show a small improvement in resolution at higher
energies~\cite{Bernabei:2012zzb} there is no updated information
available about their resolution in the low energy region.  We will
thus use \refeqn{sigmaER} to implement the detector resolution.
  
As discussed in the previous section, due to the changing WIMP
velocity distribution at Earth as the Earth orbits the Sun, there is a
small ($\sim$ 1-10\%) variation (or modulation) in the recoil rate
throughout the year \cite{Drukier:1986tm,Freese:2012xd}. The recoil
rate can be described as
\begin{equation} \label{eqn:dRdES}
  \dRdEee(\Eee,t) = S_0(\Eee) + S_m(\Eee) \cos{\omega(t-t_0)} + \ldots
\end{equation}
where $S_0$ is the average rate, $S_m$ is the modulation amplitude,
and higher order terms are often negligible (but see the discussion
below).  Note that $S_m$ can be negative, though this can only occur at
low energies.

Most direct detection experiments search for a non-modulating signal
of nuclear recoil due to dark matter scattering, \ie\ that due to
$S_0$.  These detectors have typically been smaller in fiducial volume
relative to DAMA and must be very good at background rejection.  The
strategy of the DAMA collaboration is unique in that they are looking
for the modulation amplitude, $S_m$, induced by dark matter
scattering. Their detectors have unknown, but presumably
non-modulating, backgrounds, providing a natural background rejection
strategy.  As discussed previously, the modulation amplitude is
smaller than $S_0$, however, so larger detector masses and exposure
are required.  Many modulating backgrounds have been put forward as
alternative explanations for the DAMA modulation including: detector
systematics, muons and fast neutrons produced by muons, environmental
neutrons, nuclear decays in the detector, and many others.  The
collaboration has thus far rejected each of these possibilities as
described in detail in Ref.~\cite{Bernabei:2013cfa}.

The higher order terms in \refeqn{dRdES} are usually negligible, but may
become important in some cases.  One case is in the presence of a dark
matter cold flow (\eg\ a tidal stream), where the modulation can become
very non-sinusoidal \cite{Savage:2006qr}.  Another case is for energies
corresponding to the high velocity tail of the SHM, where the modulation
becomes more sharply peaked around $t_0$.  In both cases, the amplitude
of higher order terms may become comparable to $S_m$.
However, any significant cold flow would likely change the phase of the
modulation from the SHM expected June~2, whereas the DAMA observed
modulation (May~26 $\pm$ 7~days) is consistent with the SHM.
Additionally, the DAMA modulation is also observed over a broad range of
energies ($\sim$2--5~keVee), which corresponds to a broad range of
$\vmin$ and not just the tail of the distribution.
For these reasons, a DAMA analysis can still be safely performed when
neglecting higher order terms in \refeqn{dRdES}.
In any case, the DAMA modulation amplitude determination essentially
identifies the first order Fourier expansion coefficient whether or not
higher order terms are significant.  In this case, a valid analysis can
always be performed, even if the modulation is non-sinusoidal, as long
as the predicted rates are Fourier expanded using the same phase as the
experimental determination (though ``modulation amplitude'' may lose its
meaning).

\subsection{\label{sec:Statistics}Data, Statistics, and Current Results}

\begin{figure}
  \insertfig{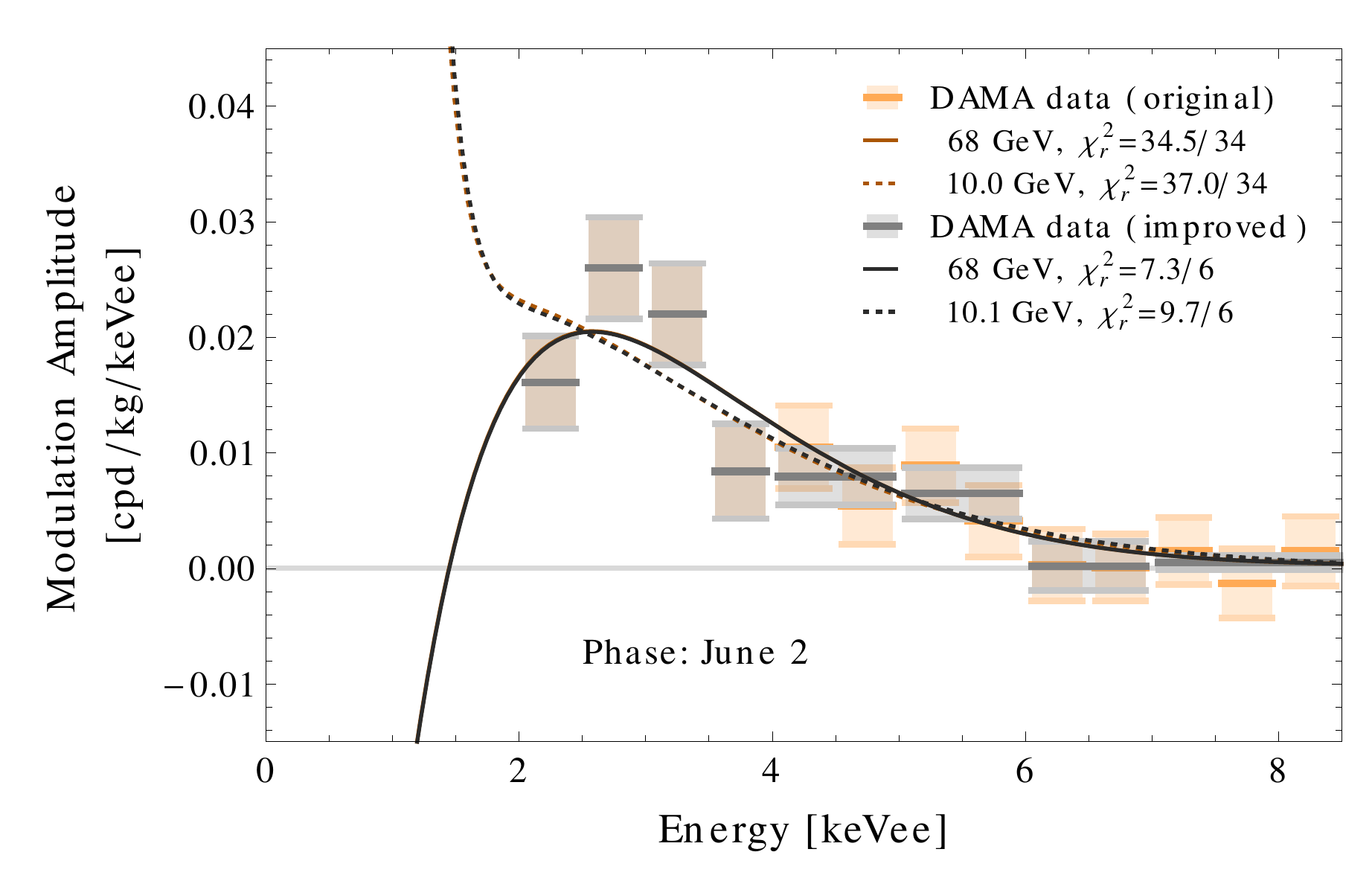}
  \caption{
  The average modulation amplitude by energy bin as measured by
  DAMA/LIBRA (orange boxes).  Though not all shown here, measurements
  extend up to 20~keVee. To improve statistical sensitivity, some of
  the original bins have been combined (gray boxes): 6 original bins
  from $4$--$7$~keVee have been combined into 3 bins with one final bin
  extending from 7-20~keVee (see the text for further discussion). The
  resulting bins are used for analyses in this paper. Also shown for
  both sets of binning are the modulation amplitude spectra for the
  WIMP mass and spin-independent (SI) cross-section that provides the
  global (solid line) and a local (dashed) chi-squared minima.
  }
  \label{fig:data}
\end{figure}

\begin{figure}
  \insertfig{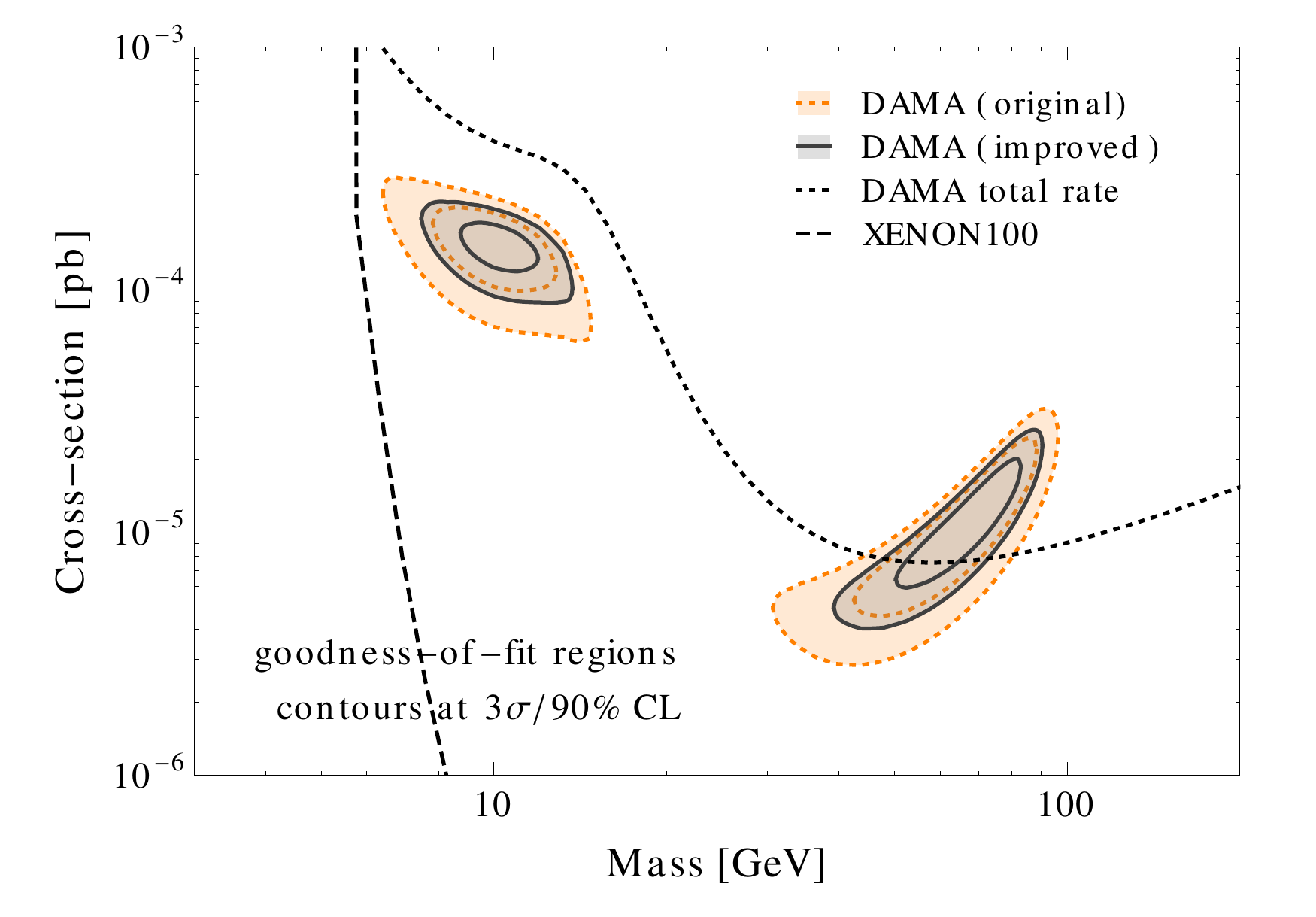}
  \caption{
    The regions of the SI cross-section vs.\ mass parameter space with
    a goodness-of-fit compatible with the DAMA data at the 90\% and
    3$\sigma$ confidence levels (CL) under the assumption of the
    standard halo model.
    Regions are shown using both the original 36 DAMA bins (orange) and
    the reduced set of 8 bins (gray) used in this analysis.
    The smaller regions in the latter case indicates the improved
    statistical sensitivity provided by combining bins; see the text
    for further discussion.
    Regions above the short-dashed curve are inconsistent at the 90\%~CL
    with the \textit{total} event rate measured by DAMA.
    Also shown for comparison is the current exclusion constraint from
    the XENON100 experiment; regions above/right of the long-dashed
    curve are inconsistent with the XENON100 result at the 90\%~CL for
    the type of coupling (SI) and halo model assumed.
    }
  \label{fig:sigmam}
\end{figure}

DAMA modulation amplitude results are presented in 36 bins, all
0.5~keVee wide, over 2--20~keVee \cite{Bernabei:2010mq}.  Measurements
over the lower part of this energy range, along with their
uncertainties, are shown in \reffig{data} (orange boxes).  An
alternative binning used in this analysis, where some of the original
bins have been combined (gray boxes), will be discussed below.  These
amplitudes have been determined assuming the SHM expected phase
($t_0$, the time of year at which Earth moves fastest relative to the
dark matter halo) of June~2; the best-fit DAMA phase is May~26 $\pm$ 7
days, consistent with this number.

Constraints on WIMP parameters and goodness-of-fit tests are based
upon the chi-square,
\begin{equation}\label{eqn:chisquare}
  \chi^2(\mchi,\bsigma)
    = \sum_k \frac{\left(S_{m,k} - S_{m,k}^T(\mchi,\bsigma)\right)^2}{\sigma_k^2} \, ,
\end{equation}
where $S_{m,k}$ is the modulation amplitude measurement in bin $k$
(averaged over energy), $\sigma_k$ is the measurement uncertainty,
$S_{m,k}^T(\mchi,\bsigma)$ is the theoretically expected amplitude for
a WIMP mass $\mchi$ and scattering cross-section(s) $\bsigma$, and the
sum is over all bins.  In most cases, we will restrict ourselves to
the WIMP parameter space containing only the mass and SI
cross-section, though we will also consider the case of SD couplings.

We test the goodness-of-fit of the WIMP framework to the data by
minimizing the $\chi^2$ over the WIMP parameter space; the resulting
$\chimin^2$ should follow a $\chi^2$ distribution with degrees of
freedom (\dof) equal to the number of data bins less the number of
parameters minimized over.  In the case of SI-only scattering, the
$\chi^2$ is minimized at $\mchi=68.3$~GeV and $\sigmapSI = 1.1 \times
10^{-5}$~pb, with $\chimin^2/\dof = 34.5/34$.  This $\chimin^2$ has a
$p$-value of 0.44 in the expected $\chi^2$ distribution and, thus, the
data are consistent with a WIMP with SI interactions.  A second, local
$\chi^2$ minimum is found at $\mchi=10.0$~GeV and $\sigmapSI = 1.5
\times 10^{-4}$~pb with $\chimin^2/\dof = 37.0/34$, also a good fit to
the data.

The global (solid orange) and local (dashed orange) best-fit spectra
are shown in \reffig{data} (both curves are obscured by black curves
of the same style).  While both spectra are a reasonable fit to the
current data, they exhibit very different behavior below the current
threshold of 2~keVee, hence the interest in a lower threshold
analysis, which would allow the two cases to be distinguished.  The
two spectra differ in that, over the current energy range of
2--20~keVee, the 68~GeV spectrum is primarily due to scattering of
WIMPs on iodine nuclei, while the 10~GeV spectrum is primarily due to
scattering on sodium nuclei.  In both cases, there are far more iodine
scatters than sodium scatters; however, in the latter case, those
iodine recoils are nearly all below threshold.  The relatively large
number of expected iodine scatters is apparent in the rapid increase
in the amplitude below 2~keVee.

Confidence regions are determined using $\Delta\chi^2 \equiv \chi^2 -
\chimin^2$.  In a two parameter space, such as the case of a WIMP with
SI-only interactions (where $\mchi$ and $\sigmapSI$ are the two
parameters), the confidence regions at a 90\%/3$\sigma$/5$\sigma$
confidence level (CL) are defined as the parameters such that
$\Delta\chi^2 \le$~4.61/11.83/28.74.

A brief aside on the use of regions in comparing experimental results is
given here.
Due to the various potential positive signals seen in some experiments
(DAMA \cite{Bernabei:2010mq}, CoGeNT \cite{Aalseth:2010vx,Aalseth:2011wp},
CRESST \cite{Angloher:2011uu}, and CDMS silicon \cite{Agnese:2013rvf})
and the lack of signal in others (XENON10 \cite{Angle:2011th},
XENON100 \cite{Aprile:2012nq} and CDMS germanium \cite{Ahmed:2009zw,
Ahmed:2010wy,Ahmed:2012vq}), the question often arises of whether these
experimental results are compatible with each other in particular WIMP
frameworks (\eg\ SI scattering assuming the SHM).  Ideally, such a question
can be answered by performing a joint likelihood analysis.  In practice,
with very different statistics used to produce constraints for different
experimental results, such an analysis proves difficult to perform.
Instead, compatibility is often approximately determined by plotting
a separately derived allowed or excluded region for each experimental
result and looking for regions in parameter space consistent with all of
the individual experimental constraints.
The question of whether there are parameters for which all experimental
results are compatible is inherently a goodness-of-fit question, of
which a $\Delta\chi^2$ confidence region alone cannot provide an answer.
The goodness-of-fit is determined separately from the $\chimin^2$ value,
and a $\Delta\chi^2$ confidence region may contain parameters that are
a very poor fit to the data if the $\chimin^2$ is poor.

An alternative method for examining compatibility is to identify
parameter ranges that give a reasonable $\chi^2$.  In the case of 36
bins, those regions at the 90\%/3$\sigma$/5$\sigma$~CL are defined by
$\chi^2 \le$~47.2/64.1/93.2; see \reffig{sigmam} for these regions in
the case of SI-only scattering.  We will refer to these as
``goodness-of-fit'' regions and the $\Delta\chi^2$ regions as
confidence regions (though, strictly speaking, the former are also
technically confidence regions).  The latter has better properties
desired in confidence regions,\footnote{%
  \textit{If} a parameter space contains the true hypothesis, the
  $\Delta\chi^2$ regions tend to be smaller than the $\chi^2$
  goodness-of-fit regions, \ie\ the former statistic is more powerful
  at rejecting incorrect parameter values.
  }
so it is preferable to use when confidence regions are wanted.  For
issues of compatibility, only the goodness-of-fit regions can be
appropriately interpreted in the manner often desired, and overlap of
such regions is a better indicator of what the ideal joint likelihood
analysis would produce.  For a brief review of statistics, see
Ref.~\cite{Beringer:1900zz}.

In addition to the modulation, DAMA measures the average total (signal
plus background) rate in their detector $\mathcal{R} = S_0+B$, where
$B$ is the background rate~\cite{Bernabei:2008yi}.  As the background
rate is unknown, DAMA cannot explicitly determine $S_0$; however, an
upper limit can be placed: $S_0 \le \mathcal{R}$.  The corresponding
constraint in the $\sigmapSI$--$\mchi$ plane is shown in
\reffig{sigmam} (black short-dashed curve); details of this constraint
shown can be found in Ref.~\cite{Savage:2008er}.  While not entirely
excluding either of the two mass regions in \reffig{sigmam}, the
constraint placed on $S_0$ does create some tension in the higher mass
region as some of this region (including the best-fit point) would
result in the signal rate alone exceeding the measured total
rate.\footnote{
  The tension in the higher mass region was somewhat stronger in the
  first (4~year \cite{Bernabei:2008yi}) DAMA/LIBRA data
  release~\cite{Chang:2008xa,Fairbairn:2008gz,Savage:2008er}.
  However, the additional two years of data added in the most recent
  release~\cite{Bernabei:2010mq} has shifted the center of the high
  mass region from $\sim$80~GeV to $\sim$70~GeV and to a lower
  cross-section, relaxing that tension somewhat.
  }
As a statistical analysis that combines both the DAMA modulation and
total rate data is rather difficult (due to the unknown background),
we shall consider only the modulation data for the remainder of the
paper, though this tension should be kept in mind.

\subsection{\label{sec:Binning}Binning}

\begin{table}
  \addtolength{\tabcolsep}{1em}
  \begin{tabular}{c@{\hspace{1em}}c}
    \hline\hline
    Energy  & Average $S_m$ \\\empty
    [keVee] & [cpd/kg/keVee]  \\
    \hline
    2.0 -  2.5 & 0.0161 $\pm$ 0.0040 \\
    2.5 -  3.0 & 0.0260 $\pm$ 0.0044 \\
    3.0 -  3.5 & 0.0220 $\pm$ 0.0044 \\
    3.5 -  4.0 & 0.0084 $\pm$ 0.0041 \\
    4.0 -  5.0 & 0.0080 $\pm$ 0.0024 \\
    5.0 -  6.0 & 0.0065 $\pm$ 0.0022 \\
    6.0 -  7.0 & 0.0002 $\pm$ 0.0021 \\
    7.0 - 20.0 & 0.0005 $\pm$ 0.0006 \\
    \hline\hline
  \end{tabular}
  \caption{
    Average modulation amplitudes observed by DAMA over the given
    energy bins.
    Some bins have been combined from the 36 bins of width 0.5~keVee
    over 2--20~keVee originally given by DAMA \cite{Bernabei:2010mq} in
    order to improve the sensitivity of statistical tests.
  }
  \label{tab:data}
\end{table}

The statistical significance of any $\chi^2$ goodness-of-fit test based
on the 36 DAMA bins will be weakened for two reasons:
(1) most of the bins are much smaller than the energy resolution of the
detector and
(2) a WIMP signal at higher energies will be negligible relative to that
at lower energies.
Both increase the number of degrees of freedom (and, thus, the level of
statistical noise in the $\chi^2$) without necessarily increasing the
signal-to-noise.
An improved choice of binning is given in \reftab{data}, along with
the corresponding modulation amplitude measurements.

To address the first issue (energy resolution), we have combined
adjacent bins that were substantially narrower than the energy
resolution at those energies.  The resulting bins still have similar or
smaller widths than the energy resolution~\cite{Bernabei:2008yh} (just
not substantially smaller).

To address the second issue (negligible signal at high energies), the
highest energy bins could all be combined into a single bin.\footnote{%
  Alternatively, the highest bins could simply be dropped.  However,
  many potential background sources of modulation would lead to a
  modulation signal over a broad range of energies; keeping one wide
  high energy bin allows the goodness-of-fit test to better exclude a
  WIMP interpretation of a signal that is due to one of these
  backgrounds.
  }
To determine the energy $E_{0}$ above which all bins should be
combined, we scanned over the the entire WIMP mass and SI cross-section
parameter space, identifying at each point the choice of $E_{0}$ that
provides the most significant deviation of the typical experimental
result from the null hypothesis (\ie\ no modulation).  We note that the
null hypothesis is used only as an example of an alternative hypothesis
(see below).
We find that
$E_{0} \le 7$~keVee is the optimal choice for any WIMP mass and
cross-section that would provide a signal with between a $5\sigma$ and
$14\sigma$ significance, with $E_{0} \le 6$~keVee preferred for signals
with a significance below $5\sigma$.  For points where the significance
is large ($\gae 14\sigma$), keeping additional bins above 7~keVee can
allow for an improved comparison with the null hypothesis as the signal
in this region, while still relatively small, becomes non-negligible.
However, when the signal is non-negligible above 7~keVee, the signal
at lower energies is extremely large, so the additional high energy
bins remain relatively unimportant.  For this reason, we choose to
combine all bins above 7~keVee into a single bin.
See \eg\ Refs.~\cite{Chang:2008xa,Savage:2008er,Kopp:2009qt} for
previous (ad-hoc) attempts to address the lack of high-energy signal
by either joining or dropping high-energy bins; our analysis here has
provided some rigor to the determination of the appropriate energy above
which the signal is statistically negligible.

A word of caution is in order here.  The alternative binning is meant
only to improve the $\chi^2$ goodness-of-fit test; other statistical
tests involving the $\chi^2$ may not benefit from the new choice of
binning.  The goodness-of-fit test is improved by increasing the test's
ability to reject a false WIMP hypothesis when the signal may be due to
any number of other plausible \textit{unknown} WIMP scenarios and/or
modulating backgrounds.\footnote{%
  There is no choice of binning for the goodness-of-fit that can
  optimally reject a false hypothesis in light of arbitrary unknown
  alternative hypotheses (one of which may be correct).
  This choice of binning aims to improve rejection of false hypotheses
  for the cases where the signal might be due to an unknown WIMP
  framework (with possibly different couplings and/or a different halo
  model than considered here), for which the modulation signal would
  still be expected to be essentially entirely below 7~keVee, or some
  modulating background that would have only a broad modulation
  spectrum above 7~keVee.  Less plausible background scenarios, such as
  one where a large modulation amplitude occurs over only a narrow
  energy window above 7~keVee, would fare worse in this case due to
  the joining of high-energy bins that would mask such a feature.
  }
For a \textit{known} alternative hypothesis, such as the background-only
(no modulation) case, a likelihood ratio analysis can be performed to
compare the WIMP and null hypotheses.
In this case, the alternative binning provides no benefit and, in fact,
can weaken the power of the ratio test, though, in practice, the
ratio test is not significantly weakened by the alternate binning here.
Likewise, parameter interval estimates based on a $\Delta\chi^2$ are
not improved by combining bins, though again, in practice, there is
little difference for the WIMP models considered here.  Since there is
little difference, to avoid confusion, we will generate confidence
regions using the $\Delta\chi^2$ based upon the 8~bins.

The new binning (gray boxes) as well as the corresponding best-fit
modulation amplitude spectra in the case of SI-only scattering (black
curves) are shown in \reffig{data}.
The best-fit SI-only scattering parameters occur at $\mchi=68.4$~GeV and
$\sigmapSI = 1.1 \times 10^{-5}$~pb with $\chimin^2/\dof = 7.3/6$,
with a second $\chi^2$ minimum occurring at $\mchi=10.1$~GeV and
$\sigmapSI = 1.5 \times 10^{-4}$~pb with $\chimin^2/\dof = 9.7/6$.
The $p$-values are 0.29 and 0.14, respectively, so both fits are
reasonably good even under the new binning.
The best-fit parameters and corresponding spectra in the two cases are
nearly identical to those found with the original binning (the curves
in \reffig{data} are almost indistinguishable), an indication that the
new binning does not significantly weaken the ability to fit WIMP
spectra to the data.

The goodness-of-fit regions derived from the new binning are shown in
\reffig{sigmam} (gray regions).  Here, the improvement in the
goodness-of-fit test is evident in the significantly reduced area of
the regions relative to the original 36 bin case.  For comparison,
the XENON100 90\%~CL exclusion limit is also shown (black long-dashed
curve); regions to the right of this curve are excluded.  While the two
experimental results are incompatible in this scenario (SI-only
scattering, SHM halo) for either set of binning, the incompatibility is
much worse with the new binning.

\subsection{\label{sec:QF}Quenching Factor}

As discussed previously, a recent measurement of the sodium quenching
factor indicated a significantly lower value with a strong energy
dependence not observed in previous studies~\cite{Collar:2013gu}.  A
smaller quenching factor would push the compatible low mass region
towards higher masses.  Using this energy-dependent quenching factor
yields a best fit mass of 11.4~GeV with $\chi^2=63.1$ for 34 degrees
of freedom $(p=1.7\times10^{-3})$. Utilizing our more optimized
binning scheme yields a best fit point at a mass of 11.4~GeV with
$\chi^2=38.1$ for 6 degrees of freedom $(p=1.1\times10^{-6})$.  We
thus find that the low mass scenario is excluded at $3\sigma$ with the
original binning, and at nearly $5\sigma$ with the optimized binning
scheme.  This example illustrates the power of our improved binning
for the goodness-of-fit test.

The exclusion of the low mass scenario can be attributed to low energy
iodine recoils. For the low mass case, the scattering is primarily
from the sodium nuclei for energies above $\sim$2~keVee, but
scattering on iodine dominates the signal at lower energies.  In fact,
it is scattering on iodine that causes the sharp rise at low energies
seen in \reffig{data}.  The fact that this steep rise must occur below
$\sim2$ keVee (given the current data) provides an effective upper
limit on the WIMP mass.  If the sodium quenching factor is decreased,
the best fit mass increases, as does the tension with the absence of
the steep rise due to iodine scattering above 2~keVee.

A measurement for the iodine quenching factor of $Q_{\rm I}=0.04$ is
also presented in Ref.~\cite{Collar:2013gu}.  This value is also
significantly lower than the $Q_{\rm I}=0.09$ found in other
studies~\cite{Bernabei:1996vj,Tretyak:2009sr}.  If both of the smaller
quenching factors from Ref.~\cite{Collar:2013gu} are used, then the
low mass scenario again becomes marginally compatible. In this case,
the best fit mass occurs at 17.8~GeV with $\chi^2=13.1$ for 6 degrees
of freedom $(p=0.041)$.

The reemergence of the low mass scenario can again be understood in
terms of the effective mass upper limit imposed by iodine scattering
at low energies.  If the iodine quenching factor is reduced, this
effectively acts to increase the upper limit on the WIMP mass due to
the absence in the data of a steep rise at low energies, allowing the
best fit point to move towards higher masses.  It is important to
note, however, the movement of the low mass region towards increasing
masses serves to further strengthen the tension with the XENON100 and
other exclusion limits.

\section{\label{sec:Models}Low-energy Models and Results}

In this section, we examine how additional low-energy modulation data
will impact constraints placed by the DAMA results.  We first examine
the details of measuring the modulation, in particular focusing on the
uncertainties in those measurements.  We then explore the simple case
of a single additional bin over 1.0--2.0~keVee, which allows us to
give a qualitative description of what additional low-energy data will
provide.  We finally turn to the likely case that data for two
additional bins, 1.0--1.5 and 1.5--2.0~keVee, will be provided by
DAMA.

\subsection{\label{sec:Measurements}Low-energy Measurements}

Of importance in identifying the impact of new low-energy modulation
measurements in DAMA is determining the approximate uncertainties in
those measurements.  Here, we discuss how the uncertainties can be
approximated.

The modulation signal must be extracted from on top of the large
average event rate $\mathcal{R} = S_0 + B$, where $B$ is the (possibly
unknown) rate per unit energy for some non-modulating background(s).
The size of the error bars in the extracted modulation amplitude should
scale with the square root of this total average rate, \ie\ a larger
rate in a detector leads to a larger uncertainty in the modulation
amplitude measurement.
For a simple estimate of this uncertainty, we use a two time bin
analysis relating the modulation amplitude $S_m$ to the average rate
in the summer minus the average rate in the winter; see
Ref.~\cite{Freese:2012xd} for details of this type of analysis.  In this
case, the error bar in a bin of size $\Delta \Eee$ will be given by
\begin{equation}
  \delta S_m
    \:=\: \frac{\pi}{2} \, \sqrt{\frac{S_0+B}{\varepsilon(\Eee)MT\Delta \Eee}}
    \:=\: \frac{\pi}{2} \, \frac{(S_0+B)}{\sqrt{N_T}}
  \label{eqn:dmError}
\end{equation}
where $MT$ is the exposure (in mass$\times$time units),
$\varepsilon(\Eee)$ is again the efficiency for detecting a nuclear
recoil, and $N_T$ is the total number of detected events.  The factor of
$\pi/2$ comes from averaging the cosine function over the half year time
bin.  More detailed analyses than our simple two bin case would lead to
a similar error estimate, though with a different leading numerical
factor than $\frac{\pi}{2}$.

In our analysis, we assume for the new low-energy bins a similar
6~year running time as with the first DAMA/LIBRA phase, yielding an
exposure of 0.87~tonne-years.  Existing measurements above 2~keVee
will have improved precision due to this additional exposure, but we
will neglect changes to the data in this energy range and use only the
current results.  Due to the $\frac{1}{\sqrt{MT}}$ scaling and the
1.17~tonne-years of existing exposure, the uncertainties on these
existing modulation measurements would be expected to drop by only
$\sim$25\% if the new exposure were included.  In the region from 1 to
2~keVee, the efficiency of the upgraded detectors is roughly
0.7~\cite{Bernabei:2012zzb}.

As the uncertainties in the low-energy measurements depend on $S_0$
and the unknown $B$, they can not be generically predicted.  However,
there are indications that the total event rate in the 1.5--2~keVee bin
just below the current 2~keVee threshold is similar to the total rate
above it (see Fig.~1 of Ref.~\cite{Bernabei:2008yi}, though the data
below 2~keVee should be treated with caution as it is below the analysis
threshold).  In this case, one can expect $\delta S_m$ in the
1.5--2~keVee bin to be similar to the $\delta S_m = 0.0040$~cpd/kg/keVee
(dru) of the existing 2.0--2.5~keVee bin measurement given the similar
exposures and total rate in \refeqn{dmError}.  Thus, for our fiducial
case, we take $\delta S_m$ to be the same 0.0040~dru in the
1.5--2.0~keVee bin.  To be conservative, we allow for the total rate
to increase in the 1.0--1.5~keVee bin and take $\delta S_m =
\sqrt{2}\times$0.0040~dru there, corresponding to a doubling in the
non-efficiency-corrected count rate.  For a single bin over 1--2~keVee,
this would correspond to $\delta S_m = 0.0035$~dru, nearly the same as
the uncertainty in the lowest existing bin, though in the single bin
analysis below we allow $\delta S_m$ to vary.

For the two additional low-energy bins analysis, we will consider a
few benchmark models consistent with a particular WIMP candidate, in
which case $S_0$ can be calculated.  For these models, if $\delta S_m$
as determined from \refeqn{dmError} is larger than our fiducial case,
even assuming $B=0$, we instead use this larger uncertainty.

\subsection{\label{sec:oneBin}Single Bin Analysis}

In this section, we examine the addition to the existing DAMA data of
a single new low-energy bin over 1--2~keVee.  Though DAMA is likely to
provide multiple narrower bins over this energy range, a single bin
will allow us to examine the qualitative effect low-energy data will
have on WIMP constraints and will also allow us to investigate how the
uncertainties in the measurements impact a modulation analysis.

The current DAMA data is consistent with two distinct regions in the
cross-section vs.\ mass plane for spin-independent scattering: a low
mass region around $\sim$10~GeV and a high mass region $\sim$70~GeV
(see Fig.~\ref{fig:sigmam}).  As can be observed from
Fig.~\ref{fig:data}, the modulation behavior of WIMPs in these two
regions would be very different in the 1--2 keVee energy range.  As
discussed in \refsec{Statistics}, in the low mass region, where higher
energy scatterings occur primarily with sodium nuclei, iodine
scatterings begin to dominate the signal for energies below 2~keVee.
This leads to the very steep rise in the spectrum just below 2~keVee.
In the high mass region, on the other hand, the scattering is
dominated by iodine over the entire energy range. This leads to the
signal turning over at around 2~keVee, corresponding to a phase
reversal where the rate is minimized in the summer rather than
maximized (see \eg\ Refs.~\cite{Green:2003yh,Lewis:2003bv} for further
discussion of the phase reversal).  The high mass scenario thus
requires a small or even negative amplitude in the 1--2~keVee energy
range opened up by the lower threshold.

\begin{figure}
  \insertfig{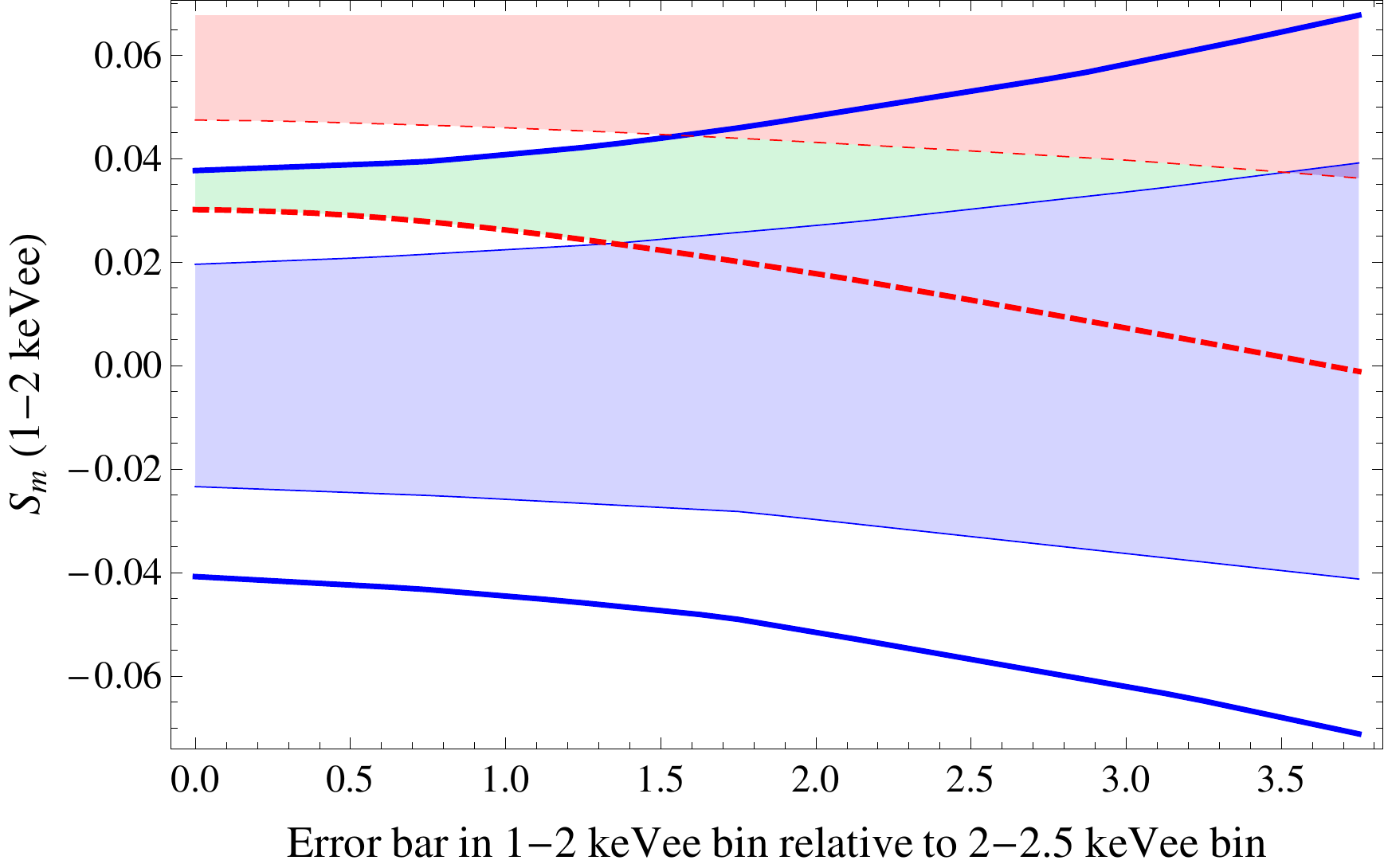}
  \caption{
    The allowed modulation amplitude in the 1--2~keVee energy range
    based on a chi-square goodness of fit.  The 90\%~CL goodness-of-fit
    region for the high mass scenario is shaded in light blue and is
    bounded by the thin, blue, solid lines and the $3\sigma$ region is
    bounded by the thick, blue, solid lines. The low mass case is shaded
    in light red and bounded by red, dashed lines.  The upper limit for
    the low mass region would be at $~\sim 0.2$
    ($\sim 0.35$)~cpd/kg/keVee for the 90\% $(3\sigma)$ regions.
    The green region indicates points excluded at 90\% (but consistent
    at $3\sigma$) for both mass ranges. 
    }
  \label{fig:amplitudeRange}
\end{figure}

The very different behavior by the two scenarios at low energy
provides the possibility that the upgraded detector could break the
degeneracy between the low mass and high mass regions.  In
\reffig{amplitudeRange} we plot the allowed modulation amplitude
assuming one new low-energy bin over 1--2~keVee based on a chi-square
goodness of fit.  The 90\%~CL goodness-of-fit region for the high mass
scenario is shaded in light blue and bounded by the thin, blue, solid
lines and the $3\sigma$ region is bounded by the thick, blue, solid
lines.  In other words, a measurement with an amplitude within the
given area is consistent with at least one set of WIMP parameters with
the mass within 25--100~GeV.  Similarly, the low mass case (5--20~GeV)
is shaded in light red and bounded by red, dashed lines.  The upper
limit for the low mass region is not indicated on the figure as it is
at a much higher scale of 0.2 and 0.35~cpd/kg/keVee for the 90\% and
$3\sigma$ regions, respectively.  The horizontal axis shows the size
of the error bar in the energy range from 1--2~keVee relative to the
size of the error bar in the current lowest energy bin (2--2.5~keVee).
This shows how the limits on the modulation amplitude improve as the
overall exposure (and thus sensitivity) increases in the lower
threshold region.

As indicated by the absence of an overlap between the red and blue
shaded regions, we find at the 90\% level that the degeneracy between
the low and high mass scenarios can be broken, even with error bars in
the lower threshold range up to 3.5 times the size of the current low
energy error bars.  If we instead expand to the $3\sigma$
goodness-of-fit region, the degeneracy can not be completely broken,
as these two regions do overlap.  This overlap region is shaded light
green in \reffig{amplitudeRange}.  In this range, all points in the
cross-section vs.\ mass plane are excluded at 90\%.  However, there
are both low and high mass WIMPs that are marginally consistent with
these measurements at the $3\sigma$ level.

\begin{figure}
  \insertfig{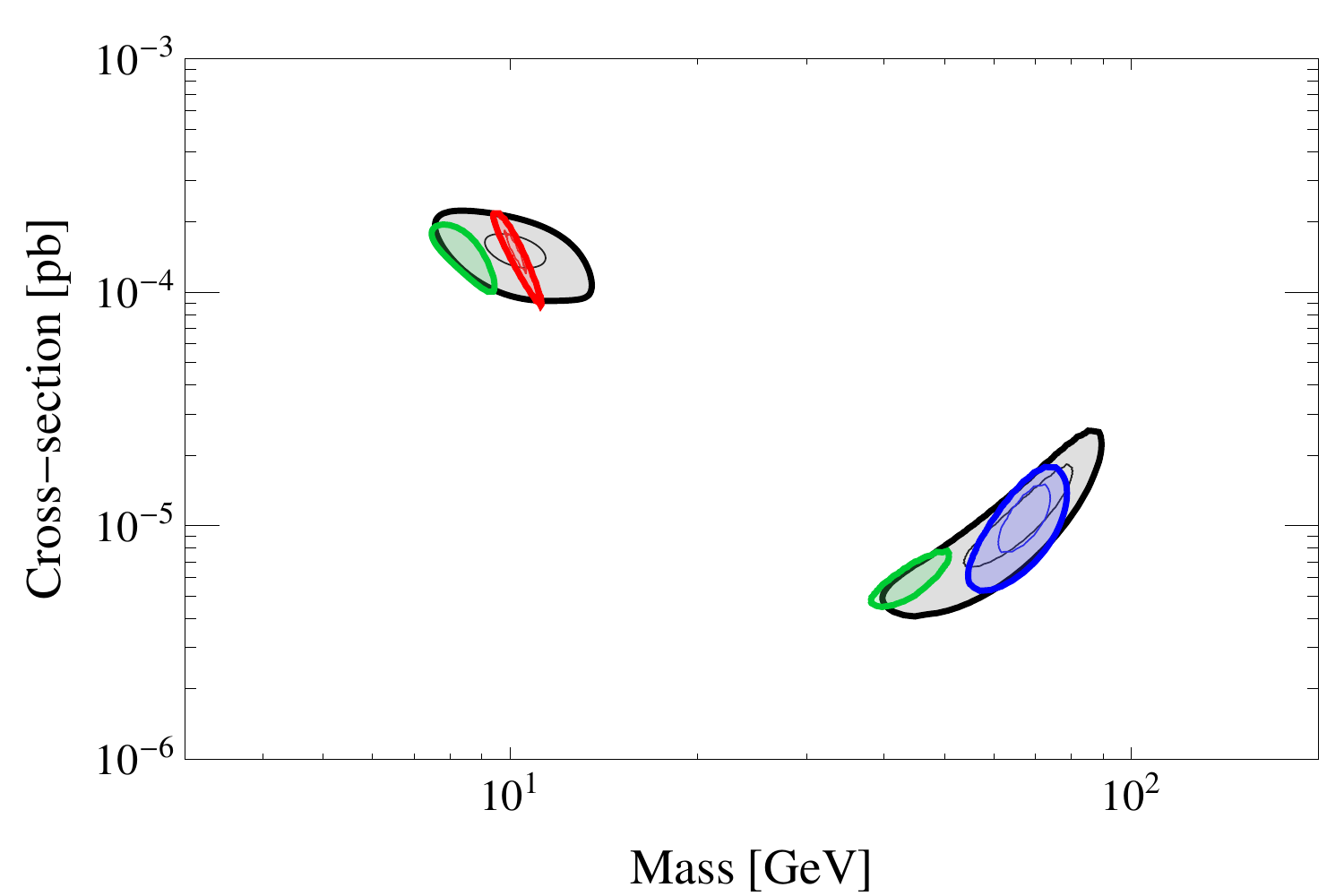}
  \caption{
    The goodness-of-fit regions for an amplitude of 0.09~cpd/kg/keVee
    (0~cpd/kg/keVee) in the energy range 1--2~keVee which is a good fit
    in the low (high) mass region is shown in red (blue). An amplitude
    measurement of 0.035~keVee shown in green, is excluded at 90\%~CL but
    marginally allowed at the $3\sigma$ level for both low and high
    masses.  The regions with current DAMA data using our more optimized
    binning are shaded grey.  In all cases, the 90\% ($3\sigma$)
    contours are shown by thin (thick) lines.
    }
  \label{fig:sigmamAllowed}
\end{figure}

Assuming dark matter is the source of DAMA's modulation, the improved
detectors with lower energy threshold will lead to tighter constraints
on the dark matter parameters.  In \reffig{sigmamAllowed}, we show the
regions allowed by the goodness-of-fit with the 90\%~CL contours
denoted by thin lines and the $3\sigma$ contours shown with thick
lines.  The grey shaded regions are the ranges using the more
optimized binning of the current DAMA data described in
\refsec{Binning} (see \reffig{sigmam}).  For all of the other cases,
we use an error bar in the 1--2~keVee energy range equal to the
current lowest energy error bar.

An amplitude measurement of 0.09~keVee (nearly 3.5 times larger than
largest current measurement) is the expected value for the low mass
best-fit point to the current DAMA data.  The resulting allowed
regions are shown in red for this low mass case.  We see that a
measurement near the best fit point in the low mass region would lead
to very small region of allowed masses and cross sections.  An
amplitude measurement of $\sim 0$~cpd/kg/keVee would be the expected
value for a point that provides a good fit to the current DAMA data in
the high mass region.  This high mass scenario is indicated in blue,
showing the modest improvement in the allowed region by lowering the
threshold.  If the amplitude is measured to be 0.035~keVee, then the
point would fall in the marginally allowed green region of
\reffig{amplitudeRange}.  This case is shown here again shaded in
light green.  As expected, all masses and cross sections for this
situation are excluded at the 90\% level, but there are allowed
regions at both low and high mass at the $3\sigma$ level.  These
regions provide a rather poor fit to the data, however, with the best
fit points having $\chi^2=18$ and $\chi^2=16$ (7 degrees of freedom)
for the high and low mass regions, respectively.

\subsection{\label{sec:twoBin}Two Bin Analysis}

\begin{table*}
  \begin{tabular}{l@{\hspace{1.5em}}c@{\hspace{1.5em}}c@{\hspace{1.5em}}c}
    \hline\hline
    & Model~1 & Model~2 & Model~3  \\
    \hline
    &\multicolumn{3}{c}{\textit{Average $S_m$} [cpd/kg/keVee]} \\
    1.0 - 1.5~keVee &  -0.0042 $\pm$ 0.0083 & 0.1484 $\pm$ 0.0057 & 0.0219 $\pm$ 0.0057 \\
    1.5 - 2.0~keVee & \ 0.0062 $\pm$ 0.0068 & 0.0258 $\pm$ 0.0040 & 0.0268 $\pm$ 0.0040 \\
    \hline
    &\multicolumn{3}{c}{\textit{Benchmark WIMP}} \\
    Mass [GeV]            & 68.4               & 10.1               & -- \\
    SI cross-section [pb] & $1.1\times10^{-5}$ & $1.5\times10^{-4}$ & -- \\
    \hline\hline
  \end{tabular}
  \caption{
    The low-energy (1--2~keVee) pseudo-data for the three models
    considered in this paper as well as the benchmark WIMP for which
    the expected spectrum is used to generate the pseudo-data
    (Models~1 and~2 only).
    All models also use the 2-20~keVee results given in \reftab{data}.
  }
  \label{tab:dataLE}
\end{table*}

\begin{table*}
  \begin{tabular}{l@{\hspace{2em}}rl@{\hspace{2em}}c@{\hspace{2em}}c@{\hspace{2em}}c}
    \hline\hline
                         & \multicolumn{2}{c}{DAMA}
      & Model~1          & Model~2          & Model~3          \\
    \hline
    \multicolumn{6}{c}{\textit{spin-independent}} \\
    \mchi{} [GeV]        & 68.4             & (10.1)
      & 67.3             & 10.2             & 50.8             \\
    \sigmapSI{} [pb]     & \scinot{1.1}{-5} & (\scinot{1.5}{-4})
      & \scinot{1.1}{-5} & \scinot{1.5}{-4} & \scinot{6.7}{-6} \\
    $\chimin^2/\dof$     & 7.3/6            & (9.7/6)
      & 8.5/8            & 10.2/8           & 14.4/8           \\
    \multicolumn{6}{c}{\textit{spin-dependent, proton-only ($\anSD=0$)}} \\
    \mchi{} [GeV]        & 10.3             & (43.7)
      & 11.0             & 3.4              & 10.0             \\
    \sigmapSD{} [pb]     & 0.60             & (0.43)
      & 0.50             & 7.1              & 0.62             \\
    $\chimin^2/\dof$     & 9.5/6            & (26.6/6)
      & 22.8/8           & 91.9/8           & 10.6/8           \\
    \multicolumn{6}{c}{\textit{spin-dependent, neutron-only ($\apSD=0$)}} \\
    \mchi{} [GeV]        & 10.0             & (52.5)
      & 58.7             & 12.3             & 47.6             \\
    \sigmanSD{} [pb]     & 84.              & (9.5)
      & 10.3             & 77.              & 9.0              \\
    $\chimin^2/\dof$     & 9.6/6            & (10.0/6)
      & 14.0/8           & 18.0/8           & 11.6/8           \\
    \multicolumn{6}{c}{\textit{spin-dependent, mixed couplings}} \\
    \mchi{} [GeV]        & 8.3              & (52.1)
      & 58.7             & 9.1              & 9.9              \\
    \apSD                & 12.0             & (0.24)
      & 0.043            & 3.7              & 1.8              \\
    \anSD                & -147.            & (-6.1)
      & -5.6             & -60.             & -4.2             \\
    $\chimin^2/\dof$     & 8.6/5            & (9.9/5)
      & 14.0/7           & 9.6/7            & 10.3/7           \\
    \multicolumn{6}{c}{\textit{spin-independent and spin-dependent}} \\
    \mchi{} [GeV]        & 67.9             & (10.4)
      & 66.9             & 9.2              & 10.0             \\
    \sigmapSI{} [pb]     & \scinot{1.1}{-5} & (0.0)
      & \scinot{1.0}{-5} & 0.0              & 0.0              \\
    \apSD                & 0.29             & (2.3)
      & 0.37             & 3.3              & 2.6              \\
    \anSD                & -0.35            & (-10.3)
      & -0.66            & -55.             & -14.             \\
    $\chimin^2/\dof$     & 7.3/4            & (9.5/4)
      & 8.4/6            & 9.6/6            & 10.3/6           \\
    \hline\hline
  \end{tabular}
  \caption{
    The best-fit mass ($\mchi$), cross-sections ($\sigmapSI$,
    $\sigmapSD$, $\sigmanSD$), and/or couplings ($\apSD$, $\anSD$)
    as well as the minimum chi-square/degrees-of-freedom
    ($\chimin^2/\dof$) for the various data sets.
    The ``DAMA'' column is for the existing DAMA data (see \reftab{data}),
    while the remaining columns are for the three models that extend
    this existing data down to lower energies (see \reftab{dataLE}).
    In the first case, parameters values at a second (local) $\chi^2$
    minimum are given in parentheses.
  }
  \label{tab:fits}
\end{table*}

\begin{figure}
  \insertfig{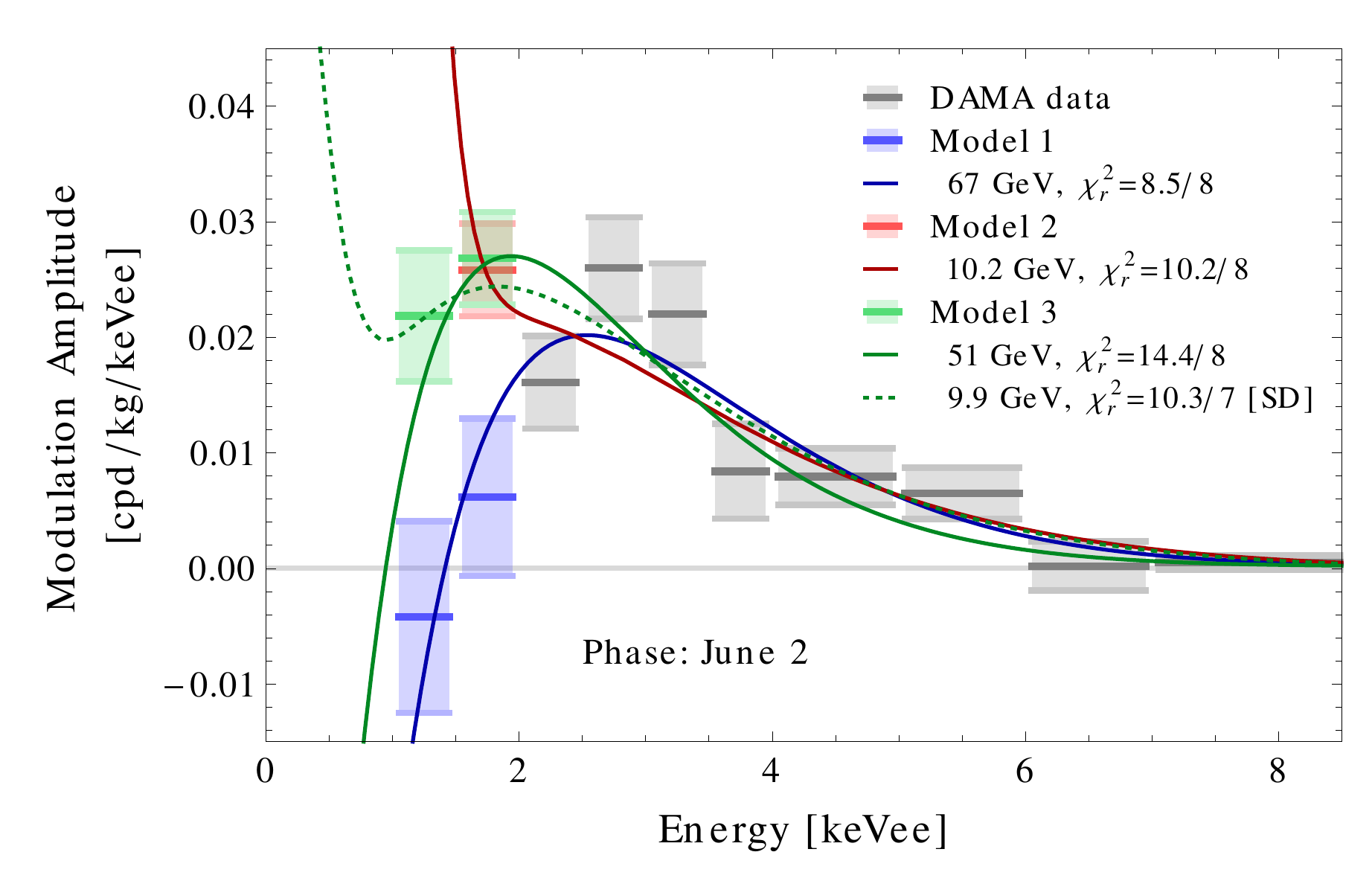}\\
  \ifmulticol{}{\hspace{0.07\textwidth}}
  \insertfig{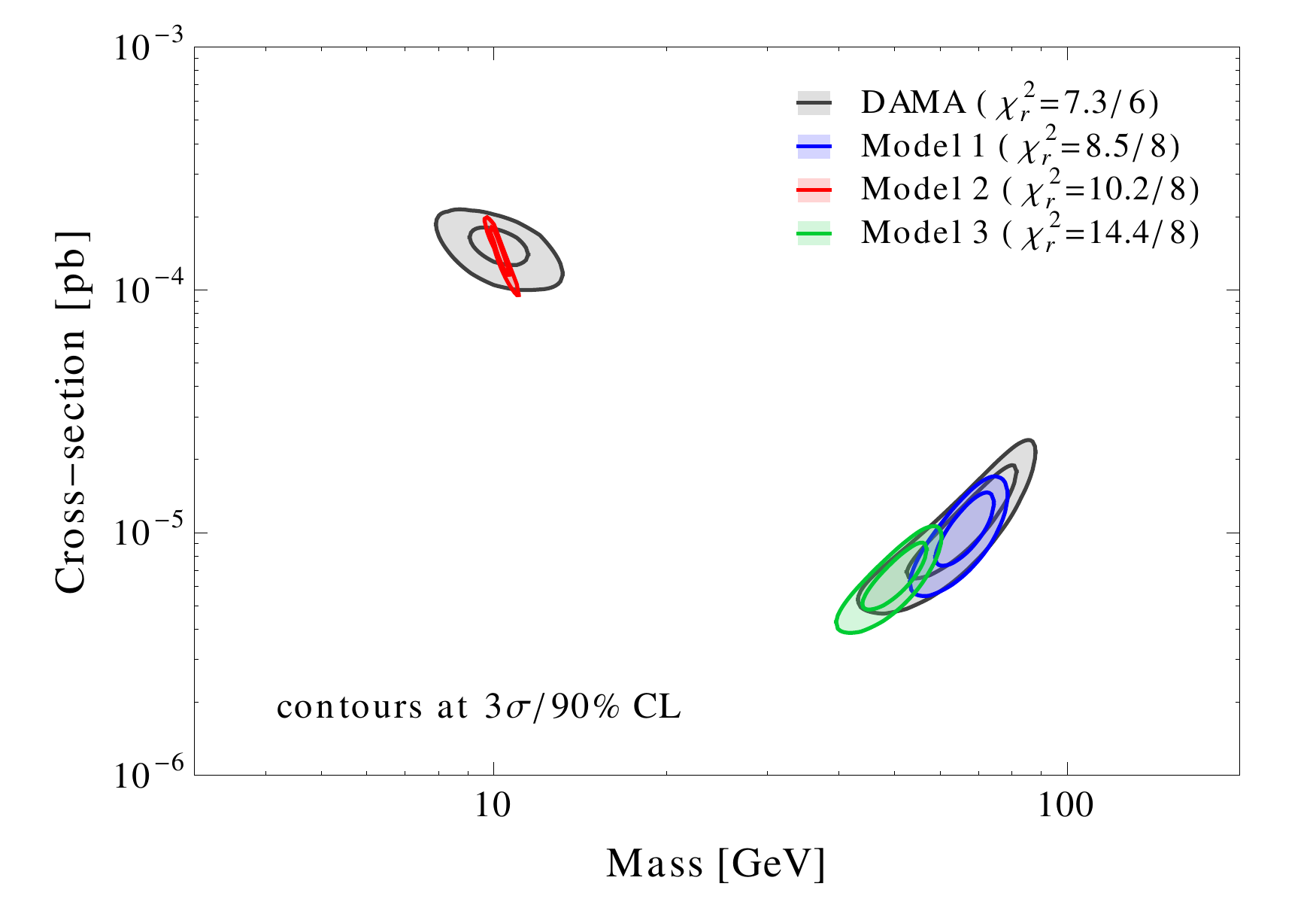}
  \caption{
    (\textit{top})
    The modulation amplitude data for the three models studied:
    pseudo-data is generated for two new low-energy bins over 1--2~keVee,
    while actual DAMA results are used for $\Eee \ge 2$~keVee.
    The first bin (1.0--1.5~keVee) of Model~2 is outside the range of
    this figure.
    Models~1 (blue) and~2 (red) are consistent with the spectra expected
    from the high- and low-mass best-fit SI points, respectively,
    while Model~3 (green) extends the measurements to lower energies
    assuming a similar modulation amplitude as observed over 2-3~keVee.
    All pseudo-data includes random fluctuations consistent with the
    estimated uncertainties.
    Also shown are the best fit SI spectra in each case (solid lines),
    as well as the the best-fit spin-dependent (SD) spectra for Model~3
    (dashed line).
    (\textit{bottom})
    The 90\% and 3$\sigma$ CL confidence regions in the SI cross-section
    vs.\ mass parameter space for the DAMA results (gray) as well as
    for the three low-energy-extended models shown in the top panel.
    The standard halo model has been assumed here.
    }
  \label{fig:LE}
\end{figure}

As discussed previously, the new threshold of the experiment is
anticipated to be 1~keVee.  If the current binning scheme is continued
into the lowered threshold region, then we would expect two additional
energy bins: 1.0--1.5~keVee and 1.5--2.0~keVee.  The two smaller bins
allows for better constraints in the WIMP parameter space relative to
the single bin discussed in the previous section.  Here, we will
examine what those constraints will be for three test cases
corresponding to three different behaviors of the low-energy
modulation spectrum.  We will additionally examine what measurements
in these two bins will be consistent with either a light or heavy
WIMP.

We define three models representative of the possible low-energy
modulation amplitude spectrum behavior: spectra consistent with a
heavy WIMP (Model~1) and a light WIMP (Model~2), and a spectrum that
assumes the modulation amplitude below the current 2~keVee threshold
remains similar to that just above it (Model~3).  For the first two
models, we choose as our benchmarks WIMPs with masses and SI
scattering cross-sections consistent with the two sets of best-fit
parameters to the existing DAMA data: the global best fit at a mass of
68~GeV (Model~1) and the local best fit at a mass of 10~GeV (Model~2).
The spectra for these two models are the ones previously shown in
\reffig{data}.  For Model~3, we assume a purely phenomenological
spectra (not based upon any particular WIMP) where $S_m$ is a constant
0.0210~cpd/kg/keVee below 2~keVee, equal to the average currently
observed amplitude over 2.0--3.5~keVee.  This is a plausible scenario
for many modulating backgrounds, where the amplitude would not
necessarily be expected to rise rapidly at low energies nor have its
phase reversed.

For each of the three models, we generate pseudo-data in the two new
low-energy bins and combine this with the existing measurements above
2~keVee.  Pseudo-data is generated by taking the true spectra as
described above and adding random fluctuations consistent with the
expected measurement uncertainties in each bin.  The uncertainties in
each bin are as described in \refsec{Measurements}, noting that the
uncertainties in the Model~1 measurements have been increased from the
fiducial case due to the larger expected $S_0$ in that scenario.  This
pseudo-data is given in \reftab{dataLE} and shown in the upper panel
of \reffig{LE}.  Note the first bin for Model~2 is outside the range
of this figure.

For each set of data, we determine the best-fit WIMP spectra assuming
SI-only couplings.\footnote{%
  The low-energy data in Models~1 and~2 were generated from the best-fit
  spectra to the existing $\ge 2$~keVee data, but that existing data is
  reused again as part of the new data set for these models.
  This results in a correlation in the data that will lead to a lower
  $\chimin^2$ than expected for a true experimental result (where all
  bins are independent), but the difference here is small and does not
  significantly alter our conclusions.
  }
Best-fit parameters are given in \reftab{fits} for each of the models
as well as the original DAMA data; best-fit parameters for the local
$\chi^2$ minima are shown in parenthesis for this last case.  These
best-fit spectra are shown as the solid lines in the upper panel of
\reffig{LE}, with the color of each line matching that of the
low-energy bins they are fitting.  For Model~1, the best-fit mass is
67~GeV, very close to the benchmark mass of 68~GeV, with
$\chimin^2/\dof = 8.5/8$ ($p = 0.39$), a good fit.  The best-fit mass
for Model~2 is 10.2~GeV, very close to the benchmark mass of 10.1~GeV,
with $\chimin^2/\dof = 10.2/8$ ($p = 0.25$), also good fit.  Model~3
has a best-fit mass of 51~GeV, but the $\chimin^2/\dof = 14.4/8$ ($p =
0.07$) indicates the fit is not so good in this case, though not poor
enough to exclude an SI-only scattering WIMP interpretation of the
modulation.

The confidence regions for each of these three data sets, again assuming
SI-only interactions, is shown in the lower panel of \reffig{LE}, with
the colors matching the corresponding bins and best-fit spectra in the
upper panel.  The confidence regions for the existing DAMA data are
shown in gray, with a region at both low and high mass showing that the
current data is insufficient to distinguish between the two masses.
The confidence regions for Model~1 (blue, centered at 70~GeV) and
Model~2 (red, centered at 10~GeV) each cover masses only near their
benchmark WIMP masses.  The additional low-energy data thus has clearly
broken the mass degeneracy.  This is as expected since the spectra, and
thus data, substantially differ below 2~keVee, even though they were
similar above 2~keVee.  For these particular models, the opposite mass
region (\eg\ the light mass region for Model~1, the high mass WIMP case)
is excluded at at least the 5$\sigma$ CL.
This suggests that, if the DAMA signal is due to a WIMP, low-energy data
will almost certainly be able to unambiguously determine if it is a
light or heavy WIMP; we shall shortly provide rigorous proof of this.
Our phenomenological Model~3 also identifies only a single mass region,
being most consistent with a heavier WIMP mass, though the low-energy
data falls somewhere between the high- and low-mass WIMP scenarios.

The previous results, based only on one set of pseudo-data each for
three specific spectral models, raises two questions.  First, given
that Model~3 data was generated from an arbitrary (non-WIMP) spectrum
but was still consistent with a WIMP interpretation, are there
low-energy measurements that could not be reasonably interpreted as
due to WIMPs?  Second, as discussed already for the single bin
analysis, assuming the data is consistent with a WIMP, will the
low-energy measurements \textit{always} distinguish between a light
and heavy WIMP mass as was the case in our three examples?

\begin{figure}
  \insertfig{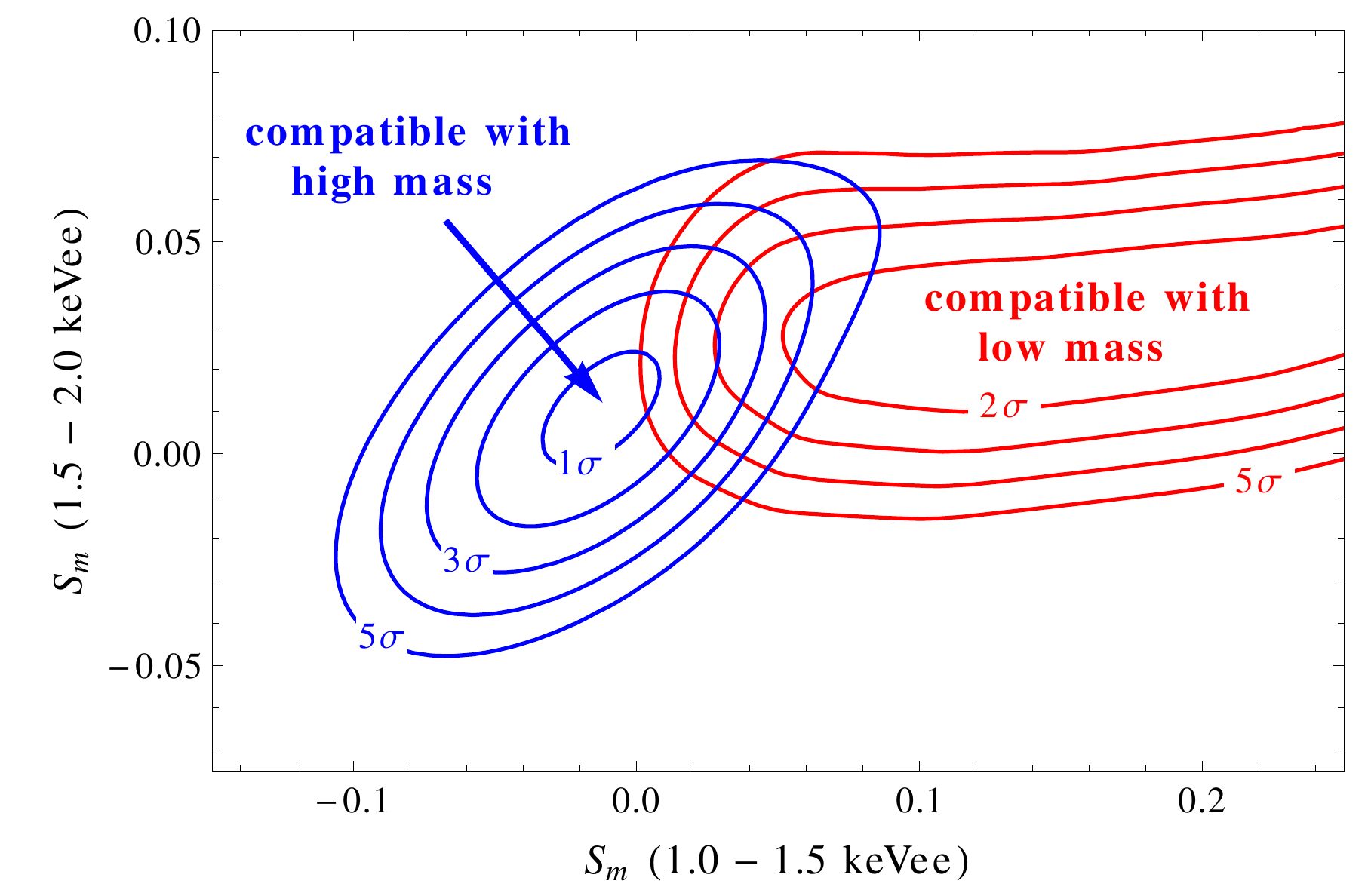}
  \caption{
    Modulation amplitude values over 1.0--1.5~keVee and 1.5--2.0~keVee
    that are compatible with light or heavy WIMPs with SI-only
    interactions.  Amplitudes are given in units of cpd/kg/keVee.
    Red (blue) contours indicate amplitudes for which at least one light
    (heavy) WIMP, defined as having a mass within 5--20~GeV
    (25--100~GeV), would provide a goodness-of-fit consistent with the
    data at the given CL.  Amplitudes outside the contours exclude
    \textit{all} light (heavy) WIMPs at at least the given CL.
    Contours are at the $1/2/3/4/5\sigma$ CL; there is no $1\sigma$
    contour for the low mass case.
    We have assumed relatively conservative error bars of 0.0083 and
    0.0068~cpd/kg/keVee for the 1.0--1.5~keVee and 1.5--2.0~keVee bins,
    respectively.
    }
  \label{fig:MassScan}
\end{figure}

\begin{figure}
  \insertfig{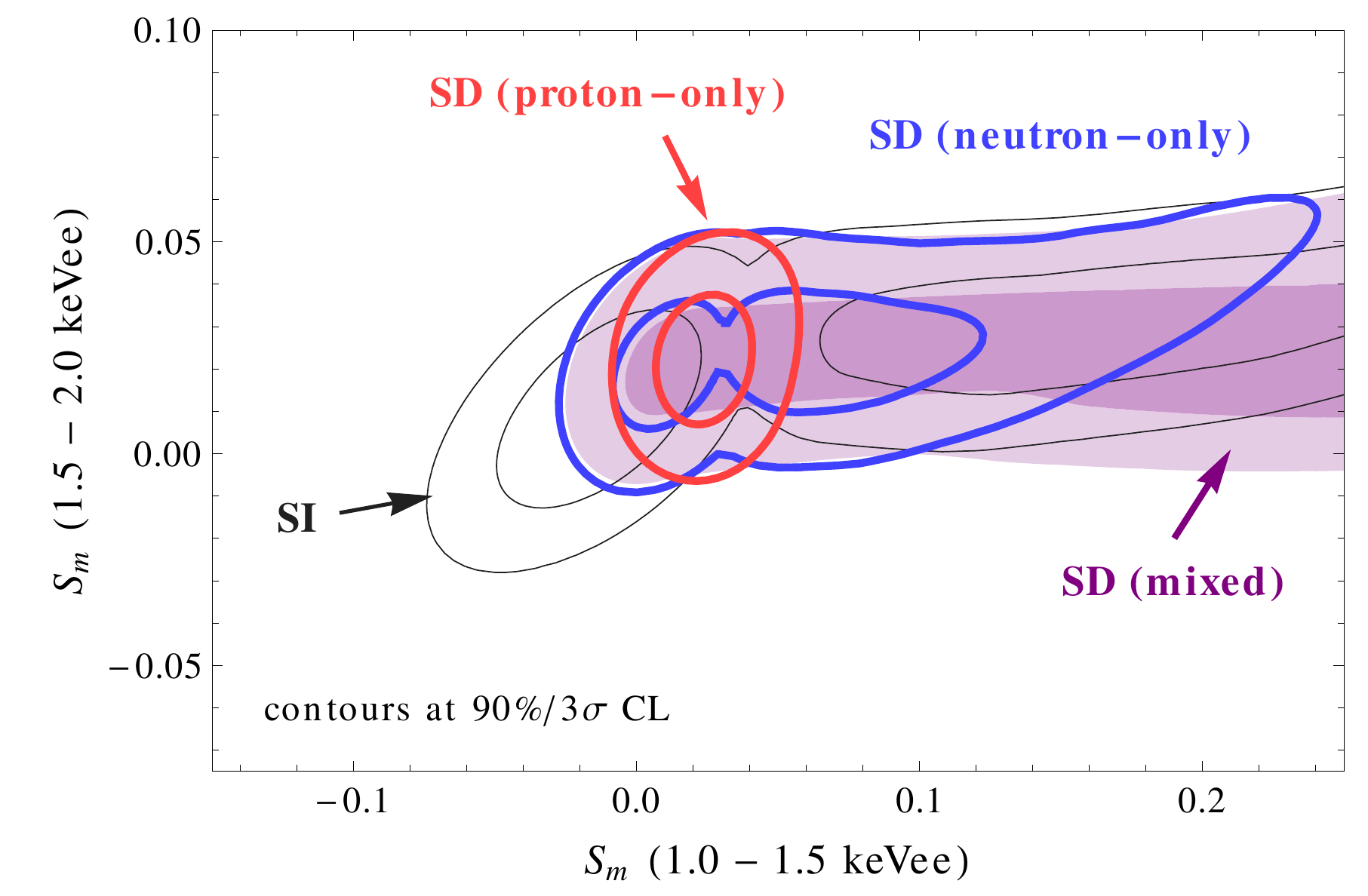}
  \caption{
    Modulation amplitude values over 1.0--1.5~keVee and 1.5--2.0~keVee
    that are compatible with different WIMP couplings at the
    90\%/3$\sigma$~CL.
    Amplitudes are given in units of cpd/kg/keVee.
    Red (blue) contours indicate amplitudes for which at least one WIMP
    with proton-only (neutron-only) spin-dependent couplings would
    provide a goodness-of-fit consistent with the data, while
    filled purple regions indicate measurements consistent with a WIMP
    of arbitrary (mixed) SD couplings.
    For comparison, the SI-allowed regions are also shown (thin black
    contours); these differ from \reffig{MassScan} in that a WIMP of
    arbitrary mass is allowed.
    }
  \label{fig:CouplingScan}
\end{figure}

To address these questions, we have scanned over all possible modulation
amplitude measurements in the two low-energy bins.  For simplicity, we
have taken the measurement uncertainties in these two bins to be equal
to that of Model~1, the model with the largest uncertainties.
For each set of amplitude measurements, we minimize the $\chi^2$ of the
full data set (two new low-energy bins plus the existing data) over
SI-only scattering WIMPs with either light (5--20~GeV) or heavy
(25--100~GeV) masses and determine the goodness-of-fit for the
$\chimin^2$.  Results are shown in \reffig{MassScan}.  For each set of
amplitude measurements within a contour, there exists at least one set
of WIMP parameters within one of the two mass ranges that is not
excluded at the given CL by the goodness-of-fit.  Outside of the
contour, \textit{all} WIMPs over the corresponding mass range would be
excluded by at least the given CL.  Contours are shown at the
$1/2/3/4/5\sigma$ CLs for both mass ranges, except there is no 1$\sigma$
contour for the light WIMP case as no set of low-energy measurements
will be consistent with a light WIMP at the 1$\sigma$ level.  The
low-mass compatible contours extend up to $\sim$0.4--0.7~cpd/kg/keVee.

The high mass contours (blue) indicate that the low-energy amplitude
measurements can be either positive or negative and still be consistent
with a heavier WIMP, though the amplitudes should not be too large.
Thus, DAMA will not necessarily see the actual phase reversal over
1--2~keVee even if the signal is due to a heavy WIMP.  On the other
hand, both measurements must be positive to be compatible with a light
WIMP at better than the 3$\sigma$ level.  For the low-mass case, the
amplitude of the first bin can be quite large, though the amplitude of
the second bin should not be substantially larger than the existing
$>2$~keVee measurements.  Comparison of the two sets of
contours shows that there is no set of measurements for which one of the
two mass ranges will not be excluded at at least the 2.6$\sigma$ level,
indicating that DAMA will \textit{always} be able to distinguish
between the two mass regions, at least at a moderate level.
This is a mild improvement over the minimum exclusion level of
2.3$\sigma$ for a single 1--2~keVee bin, though there are many two-bin
measurements that would be excluded at a far higher level than with
only the equivalent single-bin measurement.
As we have chosen to use the larger uncertainties expected
of a heavy WIMP (Model~1) rather than the smaller uncertainties more
likely for the light WIMP, the low-mass compatible regions might be
overstated and the expected distinction between high and low mass WIMPs
could be even stronger.

Up to this point, we have examined WIMPs with only SI interactions.
In \reftab{fits}, we address the possibilities of SD and mixed SI/SD
interactions by showing best-fit parameters to the existing DAMA data
and to the pseudo-data of our three models for a variety of coupling
scenarios.  For the existing DAMA data, one finds good fits in both
the low and high mass regions for all the different coupling
combinations considered (except for the high mass case with SD
proton-only scattering), with the lighter mass slightly preferred for
SD couplings and the heavier mass slightly preferred for SI and mixed
SI/SD couplings.  In other words, the current DAMA modulation results
do not prefer any particular coupling type.

The situation is remarkably different when the low-energy data in our
models is included.
For Models~1 and~2, the fits are substantially poorer (relative to the
SI case) in the cases of SD proton-only or SD neutron-only couplings,
the two cases often used to show direct detection constraints for SD
scattering.  When both SD couplings are allowed to vary, Model~2 once
again finds a set of WIMP parameters that are consistent with the data.
Thus, caution should be used when trying to compare experimental results
in the context of SD scattering in the future as the customary SD
scattering scenarios (proton-only or neutron-only) may fail to give the
full picture.  For Model~3, SD scattering gives an improvement in the
fits over the SI case.  The best-fit mixed SD coupling case gives a
spectrum as shown by the green dashed curve in the upper panel of
\reffig{LE}.  The significant decrease in the $\chi^2$ suggests that
this SD spectrum is preferred to the SI spectrum (solid green), though
quantifying the level to which SD is preferred requires a statistical
analysis beyond the scope of this paper.

While the previous examined SD scattering for only three particular
pseudo results, the general SD case is shown in \reffig{CouplingScan}.
The red (blue) contours indicate measurements for which at least one
WIMP mass and SD proton-only (neutron-only) cross-section is compatible
with the data at the 90\%/3$\sigma$~CL, again assuming the same
uncertainties as in Model~1.  The filled purple regions indicate the
same compatibility, but allowing for an arbitrary set of mixed SD couplings.
For comparison, the SI-only case is shown by the thin black contours;
these contours allow for an arbitrary WIMP mass, so they contain both
the low-mass and high-mass regions shown separately in \reffig{MassScan}.
All four 90\%~CL regions overlap in an area of the figure where the new
measurements are positive but comparable or smaller in magnitude than
the current 2--3~keVee measurement of $\sim$~0.02~cpd/kg/keVee.  For
measurements in this area, DAMA will be unable to exclude any type of
SI or SD coupling.  However, there are measurements, notably for some
negative values, for which SI scattering is consistent, but all SD
scattering is excluded; measurements here will thus be able to identify
the type of coupling.  On the other hand, measurements consistent with
SD proton-only or SD neutron-only scattering will generally also be
consistent with SI scattering.  Only the SD mixed coupling case allows
for measurements inconsistent with SI scattering, primarily at very
high amplitudes for the 1.0--1.5~keVee bin with small, positive
amplitudes for the 1.5--2.0~keVee bin.  One interesting feature of this
plot is that the SD proton-only contours encompass only a relatively
small region, meaning this particular coupling scenario could very
easily be fully excluded by future low-energy DAMA measurements.

It is clear that the lower threshold may allow DAMA to distinguish
between SI or SD interactions as the source of their modulation,
although not to a high significance level.  The reason why this may be
possible is that the relative contributions to the scattering from
sodium and iodine are different under the different couplings.  This
can be seen by comparing the Model~2 SI spectrum with the Model~3 SD
spectrum in \reffig{LE}.  These two spectra occur at nearly identical
WIMP masses (10.2 vs.\ 9.9~GeV).  The \textit{shape} of each
nucleus' contribution to the modulation spectrum is similar in both
cases, but the \textit{amplitudes} greatly differ.  Most of the events
above 2~keVee are due to sodium recoils and, thus, the spectra at those
energies are similar.  The iodine recoils occur mainly below 2~keVee
but, having fixed the appropriate cross-sections to get the sodium
scatters to match the data above 2~keVee, the total iodine scattering
rates differ by $\orderof{100}$ in the two cases, with SI interactions
giving a much larger iodine scattering rate.  The result is that the
contribution from iodine dominates over that from sodium at
$\sim$2~keVee in the SI case, but not until a much lower $\sim$1~keVee
in the SD case.

While there are regions in \reffig{CouplingScan} that indicate only one
of SI or SD couplings is compatible, in which case the data will clearly
identify which coupling type is responsible for the DAMA modulation,
there are regions where both SI and SD couplings are compatible, in
which case the type of coupling producing the signal may be less
apparent.  However, even if a measurement is compatible with both
coupling types, a hypothesis test may still indicate a strong preference
for one coupling type over another.
As neither of the SI and SD parameter spaces is a subset of the other,
the simple likelihood ratio hypothesis test is inapplicable here.
Hypothesis tests appropriate for this analysis are more difficult to
implement and computationally intensive and, as such, are not performed
here.

\Reffig{CouplingScan} shows that there are low-energy DAMA measurements
for which neither SI nor SD couplings would provide a good fit.  Thus,
new measurements could rule out the dark matter interpretation of the
DAMA modulation, at least for the standard case considered here
(non-standard WIMPs, such as velocity-dependent or momentum-dependent
WIMPs, might still be of interest; see \eg\ Refs.~\cite{Chang:2009yt,
Fitzpatrick:2012ix}).
However, one or both new low-energy bin amplitude measurements must
substantially differ from the existing $> 2$~keVee measurements.  A
background that provides a modulation amplitude that does not rapidly
change as the threshold is lowered is unlikely to produce results
incompatible with a dark matter interpretation.

Finally, we have shown that the low and high mass WIMP cases will be
clearly distinguished for SI scattering.  Does the same hold true for
SD interactions?  \Reffig{CouplingScan} does not directly answer that
question as contours are not shown separately for the two mass ranges
as was the case for SI scattering in \reffig{MassScan}.  There is some
indication from the SD neutron-only 90\%~CL contour that the two mass
possibilities may be distinguished in the SD case as well:
this contour is nearly separated into two regions, with the left region
measurements that are best fit by light WIMPs and the right region
measurements that are best fit by heavy WIMPs (here, again, light and
heavy refer to $\sim$10 and $\sim$70~GeV).  By performing separate
low and high mass scans, we find that, for SD neutron-only interactions,
at least one of the two mass ranges will be excluded at minimum at the
89\%~CL.  The SD proton-only case has \textit{already} broken the mass
degeneracy: even without new low-energy measurements, the high mass
scenario is excluded at the 4$\sigma$~CL, while a low mass WIMP is a
good fit to the existing data.  That remains unchanged with a lower
threshold, which is why the SD proton-only region is so small compared
to other couplings in \reffig{CouplingScan}.  The two masses cannot be
generically expected to be distinguished in the SD mixed coupling
case as there are measurements for which both masses provide very good
fits.  Keep in mind, however, that this is a worst case scenario.
While it is \textit{possible} to have measurements that are good fits
to both mass ranges, it is likely that the measurements will result in
a poor fit for one of the two masses.  For example, the measurements
that provide the best low (high) mass fit will exclude the high (low)
mass region at the 3.4$\sigma$ (2.9$\sigma$)~CL.
Furthermore, recall that these calculations use conservative error
estimates, so the ability to distinguish the low and high mass regions
may be stronger.

\section{\label{sec:Conclusions}Conclusions}

Recent interest in light dark matter has increased significantly due
to excess events observed in the CoGeNT, CRESST, and CDMS experiments.
The anomalies in each of these experiments have been interpreted as
evidence for spin-independent, elastic scattering of low mass WIMPs.
Although the cross section implied by the DAMA modulation is larger
than that indicated by the other experiments, it is very intriguing
that they are all pointing towards a similar mass range. Assuming
similar sized error bars, we find this low mass dark matter
interpretation of the DAMA modulation would require the amplitude in a
single energy bin from 1--2~keVee to be above 0.026~cpd/kg/keVee at
$3\sigma$, which is larger than the signal measured in any bin thus
far.  At the 90\% level, this lower limit would grow to
0.046~cpd/kg/keVee, or nearly twice the size of any of the values
measured thus far.

Additionally, even with only a single bin from 1--2~keVee, the
degeneracy between the low mass and high mass regions will be removed
at a minimum 90\% confidence level no matter what value is measured.
If the measured value is closer to one of the best-fit points in the
low or high mass region, then the exclusion of the other region would
grow considerably, to greater than $3\sigma$.  If the data is reported
using 2 bins from 1--2~keVee (as expected by the current binning
scheme), even with conservative estimates for the uncertainties, this
degeneracy will by removed at by at least the $2.6\sigma$ level.
Again, if the measured values are closer to one of the best-fit points
in the low or high mass regions, then the exclusion level of the other
region would strengthen to greater than $5\sigma$.  The ability to
discriminate so strongly between the two masses has been made possible
by selecting a binning scheme that improves the DAMA $\chi^2$
goodness-of-fit test, as discussed in \refsec{Binning}.

We used the SHM in our analysis but our results are qualitatively robust
against different assumptions regarding the velocity distribution of the
diffuse dark matter halo.
Any (plausible) halo distribution must be able fit the existing DAMA
modulation results, which is possible in two cases: a heavier WIMP
where iodine recoils dominate the 2--5~keVee signal region, and a
lighter WIMP where sodium recoils dominate this signal region.
In the latter case, a large iodine signal is unavoidably present
somewhere below the 2~keVee threshold, particularly with the $A^2$
enhancement in SI scattering.  This iodine signal should become apparent
when the DAMA threshold is lowered, providing a means to discriminate
between the two mass regions.  This qualitative picture should hold for
any reasonable background halo distribution; only the quantitative
picture should differ (though the two mass regions are unlikely to
stray far from the $\sim$10~GeV and $\sim$70~GeV regions of the SHM).

Although this work focused on the DAMA/LIBRA data, several features
are very relevant to any direct detection experiment seeking to
measure the annual modulation.  For a detector with a single element,
lowering the threshold into the region where the modulation spectrum
turns over and becomes negative would be very strong evidence for the
dark matter interpretation of the data.  It is quite difficult to
imagine a background with a spectrum that undergoes such a phase
reversal.  For a detector with more than one element, there will
always be a degeneracy at higher energies between lighter WIMPs
primarily scattering from the lighter element(s) and heavier WIMPs
primarily scattering from the heavier element(s).  This effect can be
observed with the unmodulated signal (as in the CRESST regions) or in
the modulation signal (as in the DAMA/LIBRA regions).  Once the energy
threshold is lowered far enough, however, these two scenarios make
very different predictions about the behavior of the annual modulation
signal. The annual modulation signal thus provides the opportunity to
break the degeneracy between the low and high mass regions.
Furthermore, since the relative scattering contribution from each
nucleus differs between SI and SD couplings, the modulation signal
might also discriminate between these two coupling possibilities with
a sufficiently low threshold.


\begin{acknowledgments}
  We thank the Department of Physics \& Astronomy at the University
  of Utah for support.
  CK and CS acknowledge the hospitality of the Kavli Institute for
  Theoretical Physics, which is supported in part by the National
  Science Foundation under Grant No.~NSF~PHY11-25915.
  CS also acknowledges the hospitality of the Michigan Center for
  Theoretical Physics at the University of Michigan.
\end{acknowledgments}





\end{document}